\documentclass[12pt]{iopart}

\expandafter\let\csname equation*\endcsname\relax
\expandafter\let\csname endequation*\endcsname\relax
\usepackage{amsmath}
\usepackage{amssymb}
\usepackage{graphicx} % Include figure files
\usepackage{dcolumn}  % Align table columns on decimal point
\usepackage{bm}       % bold math
\usepackage[caption=false]{subfig}
\usepackage[export]{adjustbox}
\usepackage{hyperref}
\usepackage{float}
\hypersetup{colorlinks=true, pdfstartview=FitV, linkcolor=blue, citecolor=black, 
plainpages=false, pdfpagelabels=true, urlcolor=blue}
\usepackage[all]{hypcap}
\usepackage[normalem]{ulem}

\begin{document}

\title{Parallel Temperature Interfaces in the Katz--Lebowitz--Spohn Driven Lattice Gas}

\author{Ruslan I. Mukhamadiarov, Priyanka, Uwe C. T\"auber} 
\address{Department of Physics (MC 0435) and Center for Soft Matter and Biological Physics,
		Virginia Tech, Robeson Hall, 850 West Campus Drive, Blacksburg, VA 24061, USA}
\ead{mruslani@vt.edu, pri2oct@vt.edu, tauber@vt.edu}
\vspace{10pt}
\begin{indented}
\item[] \today
\end{indented}

% =====================================================
% ABSTRACT
% ===================================================== 

\begin{abstract}
We explore a variant of the Katz--Lebowitz--Spohn (KLS) driven lattice gas in two 
dimensions, where the lattice is split into two regions that are coupled to heat baths with 
distinct temperatures. 
The geometry was arranged such that the temperature boundaries are oriented parallel to 
the external particle drive and resulting net current. 
We have explored the changes in the dynamical behavior that are induced by our choice of 
the hopping rates across the temperature boundaries. 
If these hopping rates at the interfaces satisfy particle-hole symmetry, the current difference 
across them generates a vector flow diagram akin to a vortex sheet. 
We have studied the finite-size scaling of the particle density fluctuations in both temperature 
regions, and observed that it is controlled by the respective temperature values. 
Specifically, if the colder subsystem is maintained at the KLS critical temperature, while the 
hotter subsystem's temperature is set much higher, the interface current greatly suppresses 
particle exchange between the two regions. 
As a result of the ensuing effective subsystem decoupling, strong fluctuations persist in the 
critical region, whence the particle density fluctuations scale with the KLS critical exponents. 
However, if both temperatures are set well above the critical temperature, the particle 
density fluctuations scale according to the totally asymmetric exclusion process (TASEP). 
In addition, we have measured the entropy production rate in both subsystems; it displays 
intriguing algebraic decay in the critical region, while it saturates quickly at a small but 
non-zero level in the hotter region. 
We have also considered another possible choice of the hopping rates across the temperature 
interfaces that explicitly breaks particle-hole symmetry. 
In that case the boundary rates induce a net particle flux across the interfaces that displays 
power-law behavior, until ultimately the particle exlusion constraints generate a clogging
transition to an inert state.
\end{abstract}

\vspace{2pc}

%\maketitle

% =====================================================
% I. INTRO
% ===================================================== 

\section{\label{sec:level1} Introduction}

In recent years a new interesting research avenue has emerged in non-equilibrium statistical
physics, namely studies of collective responses in spatially inhomogeneous systems. 
Whereas substantial progress has been made in understanding the origins and the often 
universal nature of cooperative behavior in systems far from equilibrium, it is still unclear 
whether it is possible to control their global collective stochastic dynamics through local 
manipulation of a finite spatial patch.
Therefore, a comprehensive characterization of spatially inhomogeneous non-equilibrium 
systems is required. 
To address this intriguing set of issues, we have chosen the Katz--Lebowitz--Spohn (KLS) 
driven lattice gas \cite{Katz:1983, Katz:1984, Schmittmann:1998, Marro:1999, 
SchmittmannBook:1995,
Caracciolo:2004}, a non-equilibrium model where particles perform biased diffusion with 
site exclusion on a lattice, and in addition experience attractive short-range Ising 
interactions. 

Interestingly, the KLS model falls into two distinct universality classes and hence displays 
very different scaling behavior depending on whether the temperature of the system that 
mediates the effectiveness of the ferromagnetic interactions is set above, or exactly at its
critical value $T_c$; and yet other dynamics in the low-temperature, ordered phase. 
The presence of these diverse types of collective motion in the same system renders the KLS 
driven lattice gas an excellent candidate for investigating the feasibility of controlling, 
through local perturbations, the universal behavior of non-equilibrium systems. 
We approach this problem by introducing a hybrid variant of the KLS model, where two 
separate sectors of the lattice are maintained at the critical and a higher temperature, 
respectively, and we investigate the resulting competition between the two distinct 
scale-invariant dynamics by analyzing the temporal evolution of this two-temperature KLS 
model.
We probe the properties of the resulting temperature interfaces, and their influence on the 
bulk scaling features of each subsystem.

To our knowledge no attempts have been taken to investigate the possibility of manipulating 
a non-equilibrium system's scale-invariant dynamics by means of local perturbations of some
external control parameter, i.e., by introducing spatial inhomogeneities.
In fact, only a few earlier studies have explored the possibility of exerting control on 
dynamical properties of non-equilibrium systems by globally varying the relevant control 
parameters \cite{Schmittmann:1994, del_Campo:2015, Karaman:2012, Priyanka:2020}. 
On the other hand, several investigations of hybrid two-temperature Ising models can be 
found in the literature \cite{Li:2012, Pleimling:2014, Colangeli:2018}. 
Some of these considered the situation when the coupling of the two temperature reservoirs 
to different parts of the lattice breaks detailed balance, driving the system away from thermal 
equilibrium \cite{Sadhu:2012, Dickman:2016, Praest:2000}; in contrast, setting up spatially 
inhomogeneous couplings in the Hamiltonian preserves detailed balance and hence 
fundamentally maintains equilibrium.

In our previous work, we studied a competition between the critical KLS and the (totally) 
asymmetric exclusion process (T)ASEP scale-invariant dynamics \cite{Janssen:1986, 
Daquila:2011, SchmittmannBook:1995} by considering a variation of the KLS driven lattice 
gas where the entire domain was split into two regions, each coupled to different 
temperature reservoirs, and with the temperature boundaries oriented perpendicular to the 
external particle drive \cite{Mukhamadiarov:2019}. 
We demonstrated that the cooler region serves as an effective transport bottleneck, impeding 
particle motion along the drive, and remarkably triggering a phase separation in the hotter 
subsystem. 
We also showed that when both temperatures exceed the critical temperature, the resulting 
density profile is well-described by the mean-field expressions derived for a (T)ASEP with 
open boundaries \cite{Krug:1991, Derrida:1992, Schutz:1995, Blythe:2007}. 
Yet if the temperature of the colder subsystem is maintained exactly at the critical value, one 
observes enhanced fluctuations, governed by the critical KLS scale-invariant dynamics.

Interestingly, the physics of the alternative geometrical arrangement for a two-temperature 
KLS model variant, namely where the temperature boundaries are oriented parallel to the 
external particle drive, appears to be very different. 
Indeed, in this paper we demonstrate that if the rates across the temperature boundaries are 
chosen to preserve the Ising $Z_2$ or particle-hole symmetry, the system remains 
(statistically) homogeneous, in the sense that no (mean) density gradients or sharp kinks 
arise. 
However, the particle current still differs in the hotter and cooler subsystems; this transport
inhomogeneity alters particle motion at the subsystems boundaries, producing a flow vector 
diagram that is reminiscent of a vortex sheet that spans the entire interface, after the net 
average particle current value is subtracted at every lattice point. 
In a quest of searching for the signatures of persisting scale-invariant dynamics, we measure
the particle density fluctuations in both subsystems, and find that the fluctuation curves scale 
with the KLS critical exponents when $T_{\rm hot} \gg T_c$ and $T_{\rm cold} = T_c$; 
but in contrast with the (T)ASEP scaling exponents when both 
$T_{\rm hot}, T_{\rm cold} \gg T_c$. 

A system's entropy production rate too is sensitive to cooperative fluctuations during 
stationarity as well as transients, and has been widely used to characterize non-equilibrium 
features through quantifying the associated probability fluxes. 
The external field in the KLS model that sets up a non-vanishing particle current necessarily
induces entropy production. 
Measuring the entropy production rate per volume in our two subsystems separately, we 
observe a power-law decay in the critical subsystem that is governed by a decay exponent 
compatible with the corresponding value recorded for the single-temperature KLS at $T_c$.
Indeed, we argue that asymptotically this temporal scaling is identical to that of the density
autocorrelation function.
Hence we conclude that critical fluctuations dominate the two-temperature KLS dynamics, if
the subsystem temperatures are set to $T_{\rm hot} \gg T_c$ and $T_{\rm cold} = T_c$.

A yet completely different scenario emerges when the hopping rates across the temperature 
interfaces explicitly violate particle-hole symmetry. 
In that perhaps more artificial situation, the system experiences a net particle flux from one 
subsystem to the other, which eventually leads to density phase separation that ultimately
terminates in a dynamically frozen state as the particles become clogged near the subsystem
boundaries. 

The outline of this paper is as follows: 
In Sec.~\ref{sec:level2} we introduce the KLS model and its two-temperature modification 
with the parallel temperature interfaces oriented along the drive direction. 
In Sec.~\ref{sec:level3} we present our results for the case when the hopping rates across 
the temperature boundaries are chosen to be $Z_2$ particle-hole symmetric. 
We conclude our work with a summary and brief discussions in Sec.~\ref{sec:level4}.
In the Appendix, we describe the consequences of choosing the hopping rates across the 
temperature boundaries such that particle-hole symmetry is manifestly broken.

% =====================================================
% II. MODEL DESCRIPTION
% ===================================================== 

\section{\label{sec:level2} Model description}

\subsection{\label{sec:sublevel21} KLS model description}

The Katz--Lebowitz--Spohn (KLS) driven lattice gas consists of $N$ particles, distributed on a 
$d$-dimensional $L_\parallel \times L_\perp^{d-1}$ lattice, where each lattice site is 
allowed to host at most a single particle, i.e., the site occupation numbers are constrained to 
$n_i = 1, 0$. 
We note that the KLS model can be equivalently described using a binary spin language, 
where instead of particles and holes the lattice is populated with up- and down-spins 
$\sigma_i = (2 n_i - 1) = \pm 1$, respectively.
In addition to the hard-core repulsion that is captured by the site exclusion restriction, these
spins or equivalently, particles and holes experience nearest-neighbor attractive Ising 
interactions.
The entire system is subject to periodic boundary conditions in all $d$ directions. 
In the KLS model particles may move to a nearest-neighbor unoccupied lattice site with a 
certain hopping rate that depends on the particle configuration on the adjacent sites, the 
temperature of the heat bath $T$ that the lattice is coupled to, and the orientation of the 
uniform external particle drive $E > 0$ \cite{Katz:1983, Katz:1984}. 

We note that the combined effect of the external applied particle drive and the periodic 
boundaries is to generate a non-vanishing particle current in the direction of the field $E$,
driving the system to a genuine non-equilibrium steady state. 
However, the existence and stability of such a stationary state necessitates that the heat 
produced by the driving field be removed from the system, providing the physical reasoning 
for coupling the lattice to an external temperature reservoir. 
As in the equilibrium Ising model, the temperature $T$ in the KLS model controls the 
efficacy of the ferromagnetic interactions.
If the total number of particles and holes in the system is chosen to be equal, i.e., the total 
particle density is set to $\rho = \frac12$, one can access the critical point in $d \geq 2$
dimensions; in two dimensions, one observes a continuous phase transition at temperatures
higher than in the corresponding equilibrium system
$T_c^{\rm KLS}(E) > T_c^{\rm KLS}(E=0) = T_c^{\rm eq}$; in the formal infinite-drive
limit, for example, $T_c^{\rm KLS}(E \to \infty) = 1.41 \, T_c^{\rm eq}$
\cite{Marro:1999}. 
In contrast to the critical point of the equilibrium Ising model, the KLS phase transition is
governed by anisotropic scaling properties induced by the external drive, and in its 
low-temperature ordered phase, the particle and hole domains assume elongated stripe-like 
shapes and are exclusively oriented along the direction of the applied field $E$ 
\cite{SchmittmannBook:1995}. 

The following set of microscopic rules constitutes the dynamics of the KLS driven lattice gas 
in $d = 2$ dimensions: 
Consider a $L_\parallel \times L_\perp$ square lattice with periodic boundary conditions, 
i.e., a two-dimensional torus, where parallel and perpendicular subscripts are relative to the 
external field $E$ orientation. 
The lattice is half-filled, with each site allowed to host one particle at most, imposing a 
restriction on possible values of the occupation number $n_i = n(x,y) = 0$ (empty site) or 
$1$ (filled site). 
The lattice dynamics is realized via nearest-neighbor hops that occur with the Markovian 
transition rates
\begin{equation} \label{eqn:KLSrates}	
	 W(\mathcal{C} \to \mathcal{C'}) \propto 
	 \exp \left(- \beta \, [H(\mathcal{C'}) - H(\mathcal{C}) + l E] \right) \ ,
\end{equation} 
where $\beta = 1 / k_{\rm B} T$, $H(\mathcal{C})$ and $H(\mathcal{C'})$ denote the 
energies of two distinct nearest-neighbor particle configurations $\mathcal{C} = \{ n_i \}$ 
with fixed total density $\rho = \sum_i n_i / L_\parallel L_\perp = \frac12$, $E$ is the 
applied particle drive strength, and $l = \{-1,0,1\}$ respectively indicates hops along, 
transverse to, and against the external bias. 
The energy of the nearest-neighbor particle configuration $\{ n_i \}$ is given by the Ising 
Hamiltonian 
\begin{equation}\label{eqn:IsingHamilt}	
	 H(\{ n_i \}) = - 4 J \sum_{\langle i,j \rangle}^N 
	 \left( n_i - \frac12 \right) \left( n_j - \frac12 \right) , 
\end{equation}
where $J > 0$ is a uniform ferromagnetic or attractive coupling stregth, and the sum is 
performed over nearest-neighbor site pairs $\langle i,j \rangle$ on the lattice.

When the external particle drive in Eq.~\eref{eqn:KLSrates} is set to zero, $E = 0$, one 
recovers the equilibrium dynamics of the Ising lattice with Kawasaki transition rates, where
in the spin language updates proceed via nearest-neighbor spin exchange processes, and the 
total magnetization is held fixed at $\sum_i \sigma_i = 0$. 
Conversely, in the limiting case of infinite particle drive strength $E \to \infty$, particle / 
hole motion in the parallel direction becomes fully biased, with all particle hops against the 
drive (and hole motion along the drive) prohibited; hence $l = \{0, 1\}$ only. 
Moreover, as the lattice temperature is increased, the nearest-neighbor interactions are 
rendered less effective; in the infinite-temperature limit $T \to \infty$, the Ising 
ferromagnetic interactions play no role at all, and the KLS model reduces to the (totally) 
asymmetric exclusion process (T)ASEP \cite{Spitzer:1970, Liggett:1985}.
Henceforth, we measure temperature $T$ in units of $J / k{\rm B}$.

\begin{table}[t]
\centering
\begin{tabular}{lclclclc}
\hline \hline \rule{0pt}{2.5ex}   
& \hspace{0.6cm} $\Delta$ & \hspace{0.9cm} $z$ & \hspace{0.9cm} $\nu$ & 
\hspace{0.9cm} $\eta$ & \hspace{0.9cm}$ \zeta$\hspace{0.4cm} \\
\hline \vspace{1mm} \rule{0pt}{2.5ex} 
Critical KLS & \hspace{0.7cm} 2 & \hspace{0.9cm} 4 & \hspace{0.9cm} 1/2 & 
\hspace{0.9cm} 0 & \hspace{0.5cm} 1/2 \\ \rule{0pt}{2.5ex} 
(T)ASEP & \hspace{0.7cm} 0 &\hspace{0.9cm} 2 & \hspace{0.9cm} -- & 
\hspace{0.9cm} 0 &\hspace{0.5cm} 1 \\
\hline \hline
\end{tabular}
\caption{Scaling exponents for the critical KLS and (T)ASEP models in two dimensions
	(omitting the logarithmic corrections for the (T)ASEP) 
	\cite{Schmittmann:1998, Leung:1986, Janssen:1986, Daquila:2012}.}
\label{tab:table}
\end{table}
As mentioned in the introduction, the generically scale-invariant non-equilibrium dynamics of 
the KLS driven lattice gas is governed by distinct sets of asymptotic scaling exponents when 
the temperature is varied. 
If the temperature of the lattice is set to $T_c(E)$, the system will be characterized by the 
KLS critical exponents \cite{Leung:1986, Janssen:1986, Daquila:2012}, with the dynamical 
correlation function obeying the following anisotropic scaling form in the steady state 
\cite{Schmittmann:1995}
\begin{equation} \label{eqn:KLSscalingform}
	C\left( x_\parallel,\vec{x}_\perp,t \right) \sim t^{-\zeta} 
	{\hat C}\left( \tau |\vec{x}_\perp|^{1 / \nu}, 
	x_\parallel / |\vec{x}_\perp|^{1 + \Delta}, t / |\vec{x}_\perp|^z \right) ,
\end{equation}
with the critical correlation length exponent $\nu$, anisotropy exponent $\Delta$, dynamic 
critical exponent $z$, $\zeta = (d + \Delta - 2 + \eta) / z$, and the Fisher exponent $\eta$ 
that characterizes the power law correlations at criticality.
Note that in two dimensions, $\zeta = \Delta / z$.
However, when the temperature is well above $T_c$, the KLS kinetics is governed by the 
(T)ASEP scaling exponents \cite{Schmittmann:1998, Mukhamadiarov:2019}. 
For reference, we present the known KLS critical exponents and (T)ASEP scaling exponents 
in two dimensions in Table~\ref{tab:table}.

\subsection{\label{sec:sublevel22} Two-temperature KLS model}

\begin{figure*}[t!]
\centering
\subfloat[\label{fig:torus}]
{\includegraphics[width=0.48\columnwidth, trim={0 0 0 0},clip]{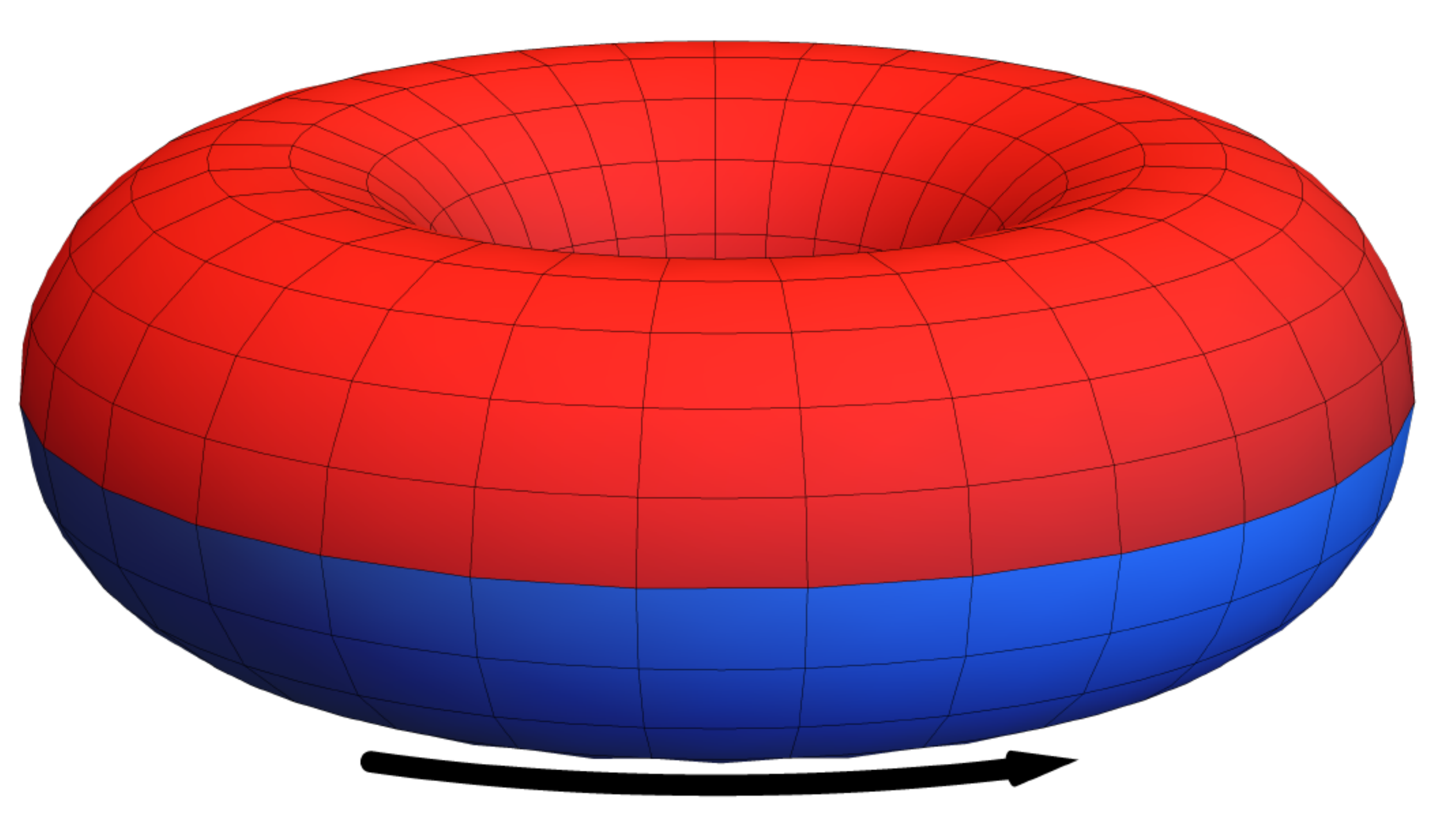}} \hfill
\subfloat[\label{fig:lattice}]
{\includegraphics[width=0.48\columnwidth, trim={0 0 0 0},clip]{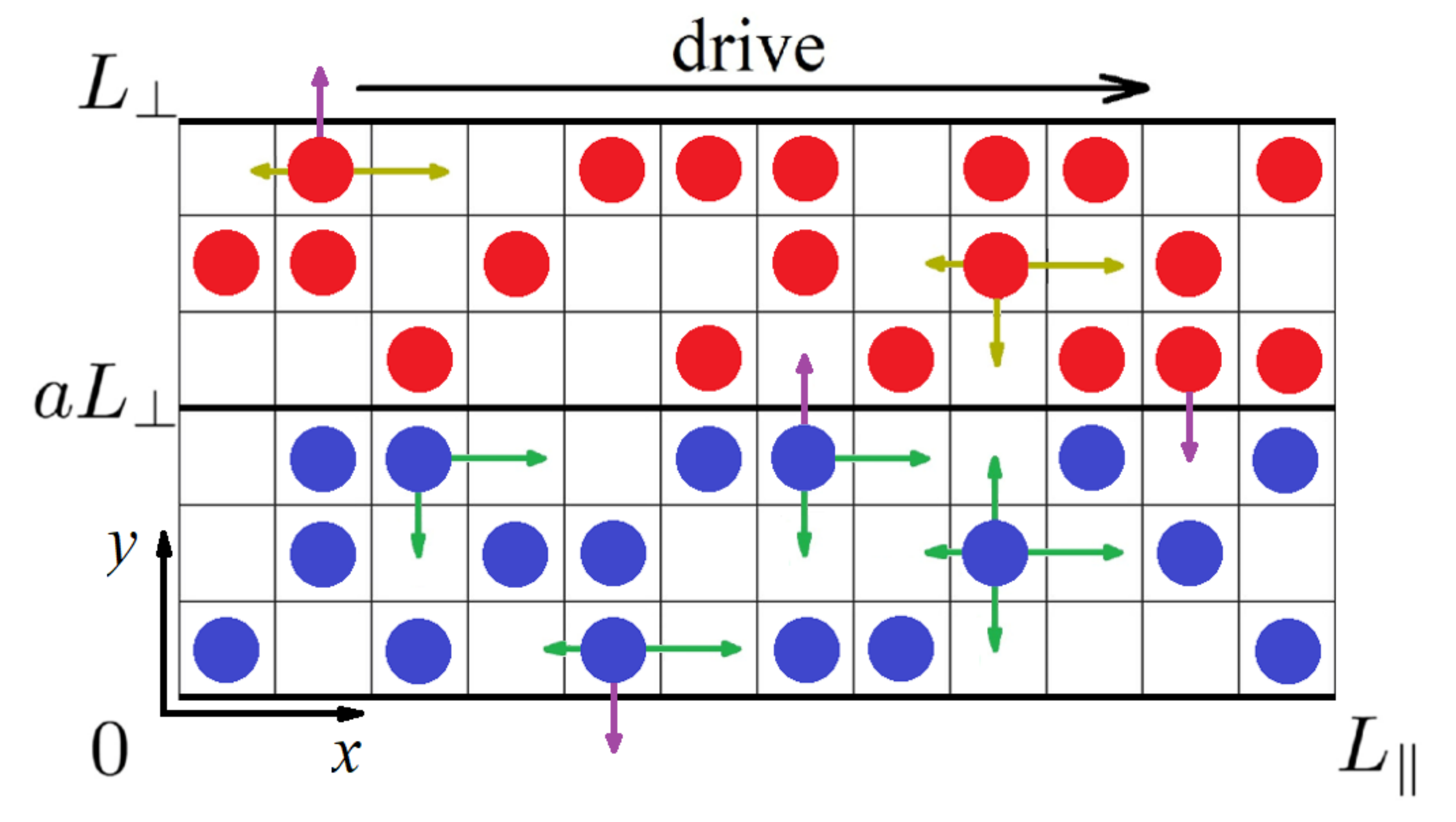}}
\caption{\label{fig:schematics}
	(a) The two-temperature driven KLS lattice gas on a ring torus. 
	The red sector of the torus is coupled to a reservoir at temperature 
	$T_{\rm hot} > T_c$, while the blue sector is coupled to a reservoir at the critical 
	temperature $T_{\rm cold} = T_c$. 
	The black arrow indicates the direction of the external particle drive. 
	(b) The two-temperature driven KLS lattice gas on an equivalent two-dimensional square 
	lattice with periodic boundary conditions. 
	The $[0, a L_{\perp})$ region of the lattice (with blue-colored particles) is maintained at 
	$T_{\rm cold} = T_c$, while the complement of the lattice is held at 
	$T_{\rm hot} > T_c$.  
	The colored arrows indicate possible hopping processes, with Metropolis rates given by 
	Eq.~\eref{eqn:KLSrates} for hops that happen within the subsystems, and by 
	Eq.~\eref{eqn:across_symm} [or Eq.~\eref{eqn:across_broken}] for hops across the
	temperature interfaces.}
\end{figure*}

Our hybrid two-temperature modification of the KLS model is comprised from two KLS 
driven lattice gases that are held at different temperatures on a ring torus, 
c.f.~Fig.~\ref{fig:torus}, and are coupled with each other by permitting particle exchange 
across the temperature boundaries. 
The two-temperature KLS model variant that we study in this paper consists of an 
$L_{\parallel} \times L_{\perp}$ rectangular square lattice with periodic boundary 
conditions, where the hotter and the cooler parts of the lattice are separated by two 
horizontal temperature interfaces that are oriented parallel to the drive: one placed at zero 
(or $L_{\perp}$) and the other at $a L_\perp$, with aspect ratio $a \in (0;1)$, as displayed 
in Fig.~\ref{fig:lattice}. 
In this schematic representation of the two-temperature KLS model, the lower part of the 
lattice $y \in [0, aL_{\perp})$ is maintained at the critical temperature 
$T_{\rm cold} = T_c$, whereas the temperature in the upper part 
$y \in [a L_{\perp}, L_{\perp})$ is set at $T_{\rm hot} > T_c$.
In this paper the subsystem size ratio is chosen to be $1:1$, which corresponds to 
$a = 0.5$. 
We shall refer to the KLS region at $T_{\rm cold} = T_c$  as the ``\textit{critical}'' 
subsystem, and to the region at $T_{\rm hot} > T_c$ as the ``\textit{hot}'' or alternatively 
the ``\textit{TASEP-like}" subsystem. 
We will also refer to the single-temperature KLS driven lattice gas as the ``\textit{standard}" 
KLS model. 

We employ the standard Metropolis algorithm with conserved Kawasaki exchange dynamics 
to simulate the evolution of the two-temperature driven KLS lattice gas on a square lattice 
with periodic boundary conditions. 
The simulation is initiated with $N = \frac12 L_\parallel L_\perp$ particles being randomly 
distributed over the lattice, and proceeds with random sequential Monte Carlo updates. 
In our simulations we set the total particle density to $\rho = \frac12$ to access the KLS 
critical point in the critical subsystem, and to avoid triggering kinetic density waves in the hot 
subsystem. 
Also, to speed up the simulation time that it takes for the system to reach its non-equilibrium 
steady state, we choose the particle drive strength to be (formally) $E = \infty$. 
A more detailed description of the algorithm can be found in our previous work 
\cite{Mukhamadiarov:2019}. 

Eq.~\eref{eqn:KLSrates} prescribes the hopping rates in the bulk of each subsystem.
However, one has to decide how to handle particle hops across the temperature interfaces. 
We impose the constraints that this choice should preserve (i) detailed balance in order to not 
trigger net particle currents between the subsystems, and (ii) the fundamental KLS 
particle-hole or Ising $Z_2$ symmetry; therefore it must account for the nearest-neighbor 
ferromagnetic interactions, too. 
As we demonstrate in the Appendix, violating the detailed-balance condition by breaking the
particle-hole or Ising $Z_2$ symmetry produces a net particle flux across both temperature 
boundaries. 
Thus, it is paramount for our study to choose a proper mathematical form for the interface 
hopping rates that couple the two distinct temperature regions. 
Considering the two aforementioned properties, we propose the following particle-hole 
symmetric rate prescription for hops across the temperature interfaces:
\begin{equation} \label{eqn:across_symm}
     W(\mathcal{C} \to \mathcal{C'}; T_1 \to T_2) \propto \exp \left[
     - \left( \frac{H^p(\mathcal{C'})}{T_2} + \frac{H^h(\mathcal{C'})}{T_1} \right) 
    + \left( \frac{H^p(\mathcal{C})}{T_1} + \frac{H^h(\mathcal{C})}{T_2} \right) \right] ,
\end{equation}
where $T_1$ is the temperature of the subsystem that the particle tries to leave, and $T_2$ 
is the temperature of the subsystem that the particle attempts to enter. 
Here the energy function $H(\mathcal{C})$ from Eq.~\eref{eqn:IsingHamilt} is split into 
two parts, $H(\mathcal{C}) = H^p(\mathcal{C}) + H^h(\mathcal{C})$, with 
$H^p(\mathcal{C}, \{ n_i=1 \}) = - 2 J \sum_{\langle i,j \rangle}^N \left( n_j - \frac12 
\right)$ representing the part of the energy sum over the occupied (particle) sites' ($i$) 
nearest neighbors $j$, while $H^h(\mathcal{C}, \{ n_i=0 \}) = 2 J \sum_{\langle i,j
\rangle}^N \left( n_j - \frac12 \right)$ is the complementary contribution to the Hamiltonian 
that extends over the holes' neighbors. 

It is straightforward to ascertain that the interface hopping rates \eref{eqn:across_symm} 
preserve particle-hole (Ising $Z_2$) symmetry, i.e., upon replacing all particles with holes
and vice versa in the configurations of interest, the transition rates from the resulting new 
configurations $\mathcal{C}$ to $\mathcal{C'}$ will be equal to the original ones: 
$\sum_{\mathcal{C}} W(\mathcal{C} \to \mathcal{C'}; T_1 \to T_2) = 
\sum_{\mathcal{C}} W(\mathcal{C} \to \mathcal{C'}; T_2 \to T_1)$. 
As we show in the following section, this equality ensures that our choice of the hopping 
rates across the temperature boundaries does not give rise to a net particle current between 
the two subsystems, which in turn would cause density gradients in the transverse direction. 
In the appendix, we briefly discuss the consequences of selecting interface hopping rates that 
explicitly violate particle-hole symmetry.

We remark that coupling the lattice to two temperature reservoirs in this manner differs from 
a situation when the two sectors of the lattice are maintained at the same temperature, but 
have different interaction strengths $J$ in Eq.~\eref{eqn:IsingHamilt}. 
In the former situation hops across the temperature interface violate detailed balance, while
it remains intact in the latter scenario, which represents a spatially inhomogeneous Ising 
model.

% =====================================================
% III. Results
% ===================================================== 

\section{\label{sec:level3} Simulation results}

\begin{figure*}
\centering
\includegraphics[width=0.48\columnwidth]{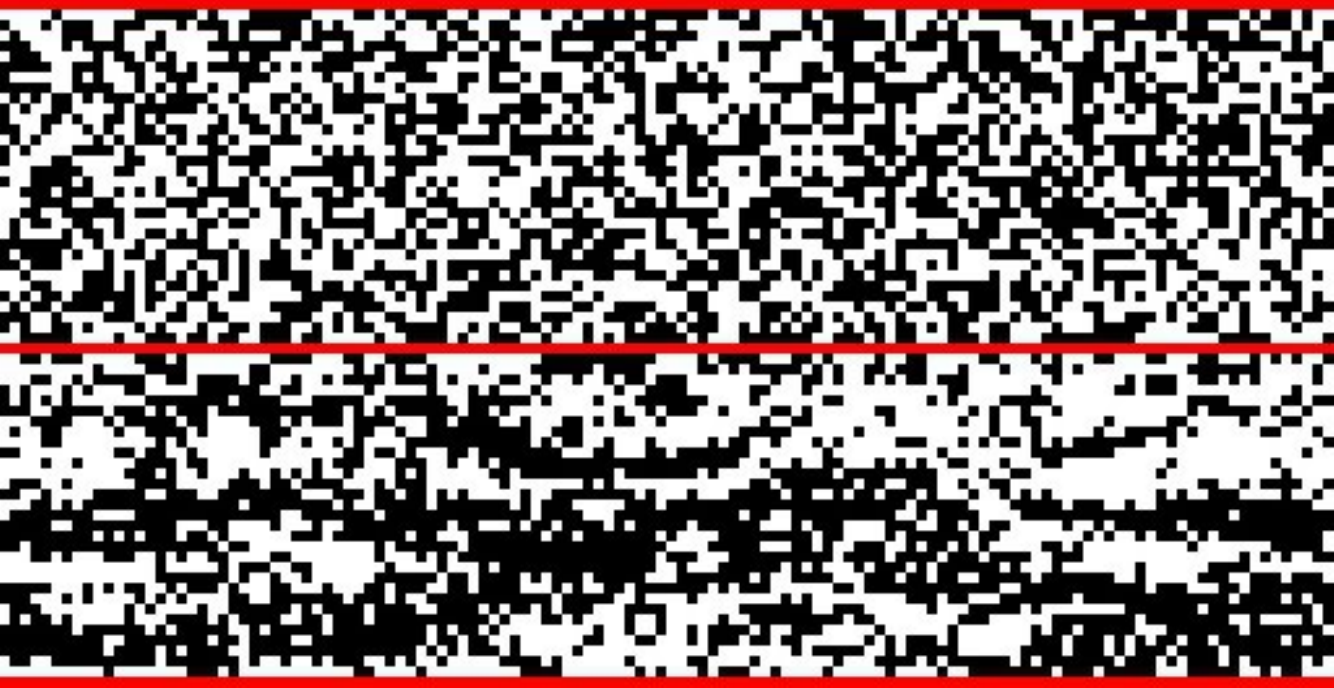}
\caption{\label{fig:snapshot}
	Simulation snapshot of the two-temperature KLS driven lattice gas with the	
	temperature boundaries aligned parallel to the drive in its steady state. 
	The system size is $L_\parallel =  128$ and $L_\perp = 64$, with subsystem ratio 
	$1:1$ ($a = 0.5$). 
	The red lines at $y = L_\perp / 2$ and $y = 0 = L_\perp$ indicate the positions of the
	temperature interfaces.  
	On this snapshot, the upper part of the lattice is coupled to the $T = 10.0$ temperature 
	reservoir, and the bottom part to the $T = T_c^{128x32} = 0.782$ reservoir.}    
\end{figure*}
Running the simulations for the two-temperature KLS model with the particle-hole symmetric
hopping rates across the temperature boundaries \eref{eqn:across_symm}, we have found 
that the system remains homogeneous; the average particle density remains $\frac12$ along 
each slice in the transverse direction, and no net particle flux is present perpendicular to the 
drive. 
As depicted in the stationary-state snapshot Fig.~\ref{fig:snapshot}, elongated clusters form 
in the critical subsystem along the direction of the drive, while the hotter region remains 
disordered at all times. 
On first glance the two subsystems seem to evolve as if they are not coupled at all. 
However, when we checked the initial aging scaling in the two subsystems separately, we 
found that scaling in each subsystem persists only for a short period of time. 
After quenching both subsystems from the random initial state to their different assigned 
temperatures, we measured the two-time autocorrelation function 
$C(\vec{x}=0; t ,t_w) = \langle n(0,t) n(0,t_w) \rangle - \rho^2$, where $t_w < t$ is often 
termed waiting time. 
Using the simple aging scaling form 
$C(\vec{x}=0; t, t_w) = t_w^{- \zeta} {\hat C}(t / t_w)$, we looked whether the two-time 
autocorrelation function in the critical and hot subsystem scaled with the KLS critical 
exponents and with the (T)ASEP aging scaling exponents, respectively 
\cite{Mukhamadiarov:2019, Daquila:2012, Daquila:2011}. 
As over time the dynamical coupling between the subsystems becomes manifest, the simple
aging scaling of the two-time autocorrelation function breaks down in both regions, 
indicating that the mutual interaction between the subsystems alters the scale-invariant 
dynamics in both temperature regions, similarly to what we had previously observed in the 
two-temperature KLS driven lattice gas with transverse temperature boundaries
\cite{Mukhamadiarov:2019}.

\subsection{\label{sec:sublevel31} Current profile}

\begin{figure*}[t!]
\centering
\includegraphics[width=0.67\columnwidth]{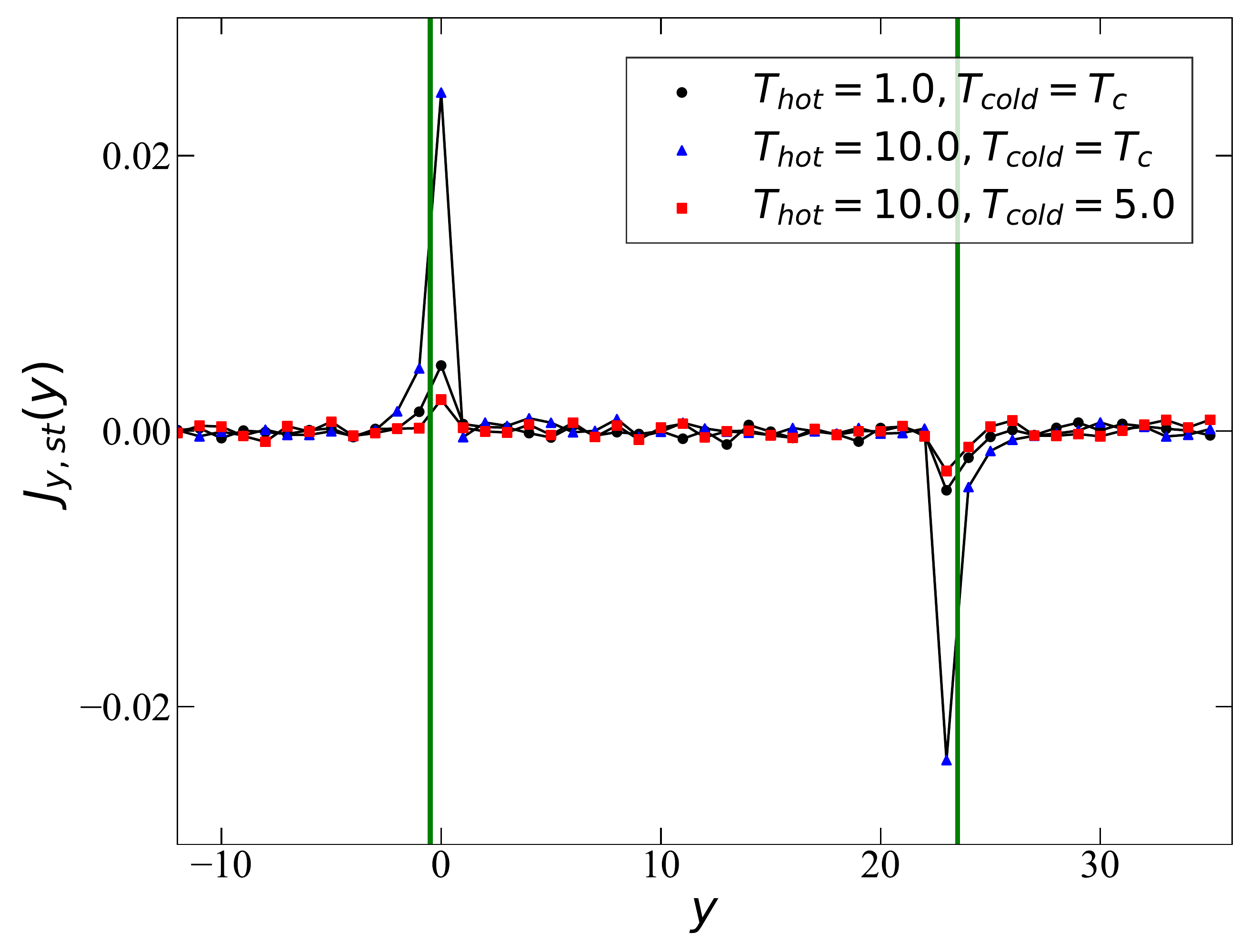}
\caption{\label{fig:curprof}
	The current profile of the two-temperature KLS driven lattice gas with dimensions 
	$L_\parallel =  54$, $L_\perp = 48$ and $1:1$ subsystem size ratio for different 	
	temperatures of the critical and hot subsystems; the value of the KLS critical temperature 
	for the $54x24$ system is $T_c^{54x24} = 0.773$.
	The hot subsystem is located in the middle of the graph; the temperature boundaries 
	are indicated by green lines. 
	The data were averaged over $10,000$ independent realizations.}
\end{figure*}
To understand how the subsystems' coupling modifies the dynamics in each temperature 
region, we first analyzed the current profile of the transverse current in the $y$ direction.
Our simulations yield that the transverse particle current remains zero in the bulk of both 
subsystems, but exhibits spikes at the hot subsystem boundaries, c.f.~Fig.~\ref{fig:curprof}. 
These perpendicular current peaks originate from the fact that on average the hopping rates 
in the hot subsystem $\langle W(T_{\rm hot}) \rangle$, Eq.~\eref{eqn:KLSrates}, exceed
the mean hopping rate value across the temperature boundaries 
$\langle W(T_{\rm hot} \to T_{\rm cold}) \rangle$, Eq.~\eref{eqn:across_symm}. 
Consequently, particles are likely to be screened from the hot region's boundaries as they 
approach it.
As the difference between the hot subsystem's bulk hopping rates and the hopping rates 
across the temperature interfaces increases, the current peaks at the temperature boundaries 
become larger and broader, reaching their maximum value for $T_{\rm hot} = \infty$ and 
$T_{\rm cold} = T_c$. 
(We do not consider temperatures $T < T_c$ in this work). 

\begin{figure*}[t!]
\centering
\subfloat[\label{fig:curvecplot_1}]{\includegraphics[width=0.49\columnwidth, 
	trim={0 0 0 0},clip]{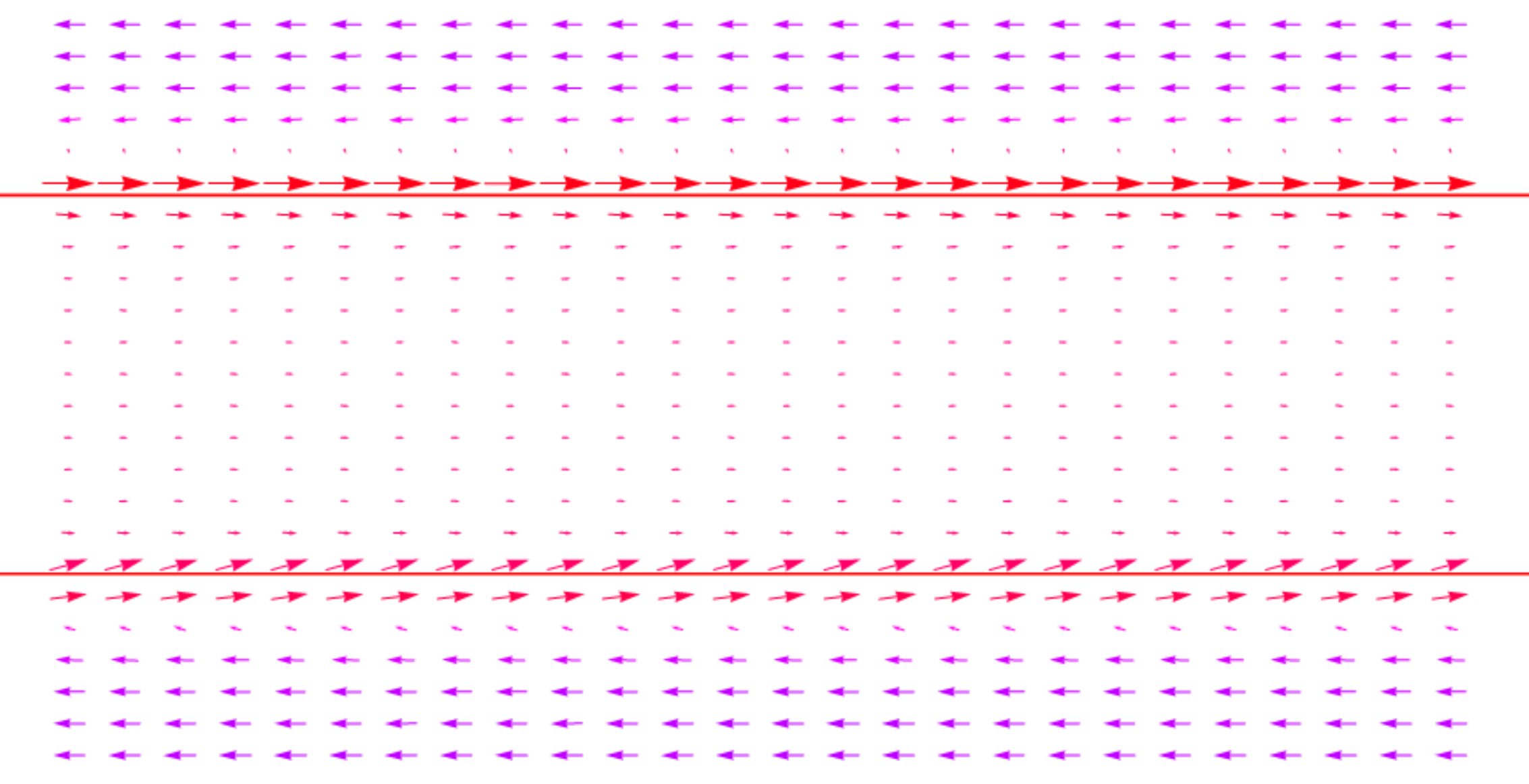}} \hfill
\subfloat[\label{fig:curvecplot_10}]{\includegraphics[width=0.49\columnwidth, 
	trim={0 0 0 0},clip]{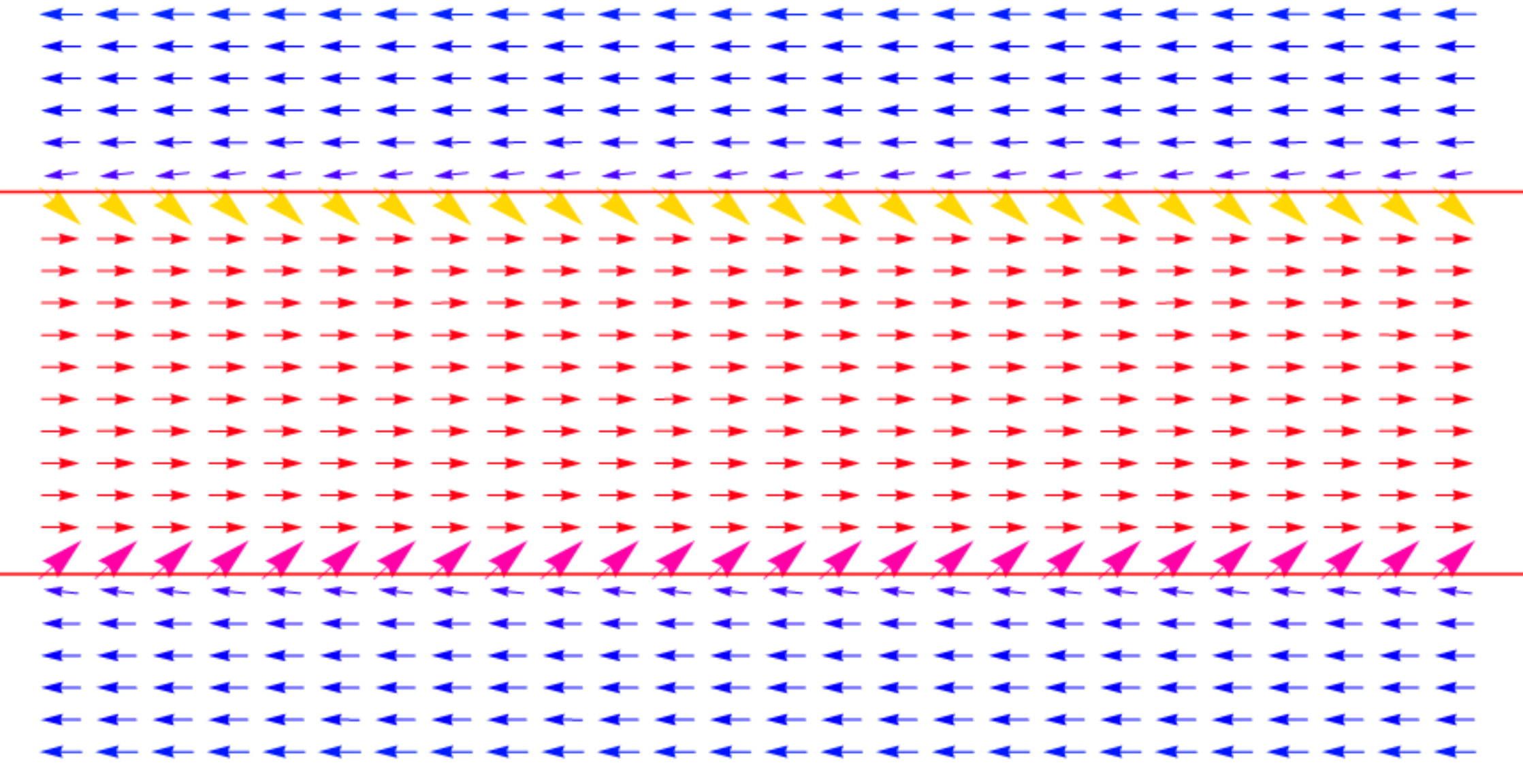}}
\caption{\label{fig:curvecplot} 
	The coarse-grained current vector plot for the two-temperature KLS model in the steady 
	state after the average current due to the external particle drive has been subtracted. 
	The temperatures of the subsystems here were chosen as follows: 
	(a) $T_{\rm hot} = 1.0$, $T_{\rm cold} = T_c^{54x24} = 0.773$; 
	(b) $T_{\rm hot} = 10.0$, $T_{\rm cold} = T_c^{54x24} = 0.773$. 
	The hot temperature region is located in the middle of the plot, with the red lines 
	indicating the positions of the temperature boundaries. 
	The system size is $L_\parallel =  54$, $L_\perp = 48$, with $1:1$ subsystem size 
	ratio. 
	The data were averaged over $10,000$ Monte Carlo steops and $1,000$ independent 
	realizations.}
\end{figure*}

We have performed a more detailed analysis of the non-trivial structure of the local particle 
transport in the two-temperature KLS driven lattice gas by studying the current vector field in
the hybrid lattice. 
To render boundary effects more apparent, we subtracted the average net current value from 
each lattice point and thus obtained the coarse-grained current vector plot shown in 
Fig.~\ref{fig:curvecplot}, where each arrow represents the sum of four current vectors from 
the adjacent four lattice sites. 
The resulting current vector patterns for the two-temperature KLS model with parallel 
temperature interfaces resemble two vortex sheets located at the subsystem boundaries that 
span the entire system.
Such vortex sheets are known to appear in fluid-mechanical systems that display velocity 
discontinuities \cite{Anderson:1987}. 
In our two-temperature KLS model, this intriguing boundary feature emerges from the 
parallel particle current difference in both subsystems:
The particle current along the drive in the critical subsystem is lower than in the hot region,
since the formation of stripe-like clusters in the critical subsystem impedes transport 
\cite{Mukhamadiarov:2019}.
The strength of these vortex sheets is sensitively controlled by the temperature gradient 
between the two regions. 
As is shown in Fig.~\ref{fig:curvecplot_10}, when the temperature of the hot subsystem is
increased, the strength of the vortex sheet grows accordingly, reaching its maximum for 
$T_{\rm hot} \to \infty$ and $T_{\rm cold} = T_c$, and vanishing when 
$T_{\rm hot} = T_{\rm cold}$, or when both temperatures 
$T_{\rm hot}, T_{\rm cold} \gg T_c$.

\subsection{\label{sec:sublevel32} Density fluctuations}

Looking for signatures of the scale-invariant dynamics in the two-temperature KLS model, 
we found that the particle density fluctuations in each subsystem exhibit intriguing 
non-trivial scaling behavior. 
While the total number of particles in the entire system of course stays constant, and the 
average density in each subsystem remains one-half, the hotter and cooler regions may 
exchange particles, whence the total number of particles in each subsystem fluctuates. 
We used the following expression to define the integrated particle density fluctuations (per 
volume) in the entire critical subsystem:
\begin{equation} \label{eqn:denfluct}
    \left( \Delta \rho(t) \right)^2 = \Bigg\langle \bigg( 
    \frac{1}{L_{\parallel} \cdot a L_{\perp}} \sum_{i,j}^{L_{\parallel}, a L_{\perp}} 
    n_{ij}(t) \bigg)^2 \Bigg\rangle - \Bigg\langle \frac{1}{L_{\parallel} \cdot a L_{\perp}} 
    \sum_{i,j}^{L_{\parallel}, a L_{\perp}} n_{ij}(t) \Bigg\rangle^2 ,
\end{equation}
where $L_{\parallel} \cdot a L_{\perp}$ is the size of the critical subsystem, and the 
averaging is performed over different, independent simulation runs. 
To obtain the particle density fluctuations in the hot subsystem, one needs to simply replace 
$a L_{\perp}$ everywhere in Eq.~\ref{eqn:denfluct} with $(1-a) L_{\perp}$.

\begin{figure*}[t!]
\centering
\subfloat[\label{fig:denfluct_TASEP}]{\includegraphics[width=0.49\columnwidth, 
	trim={0 0 0 0},clip]{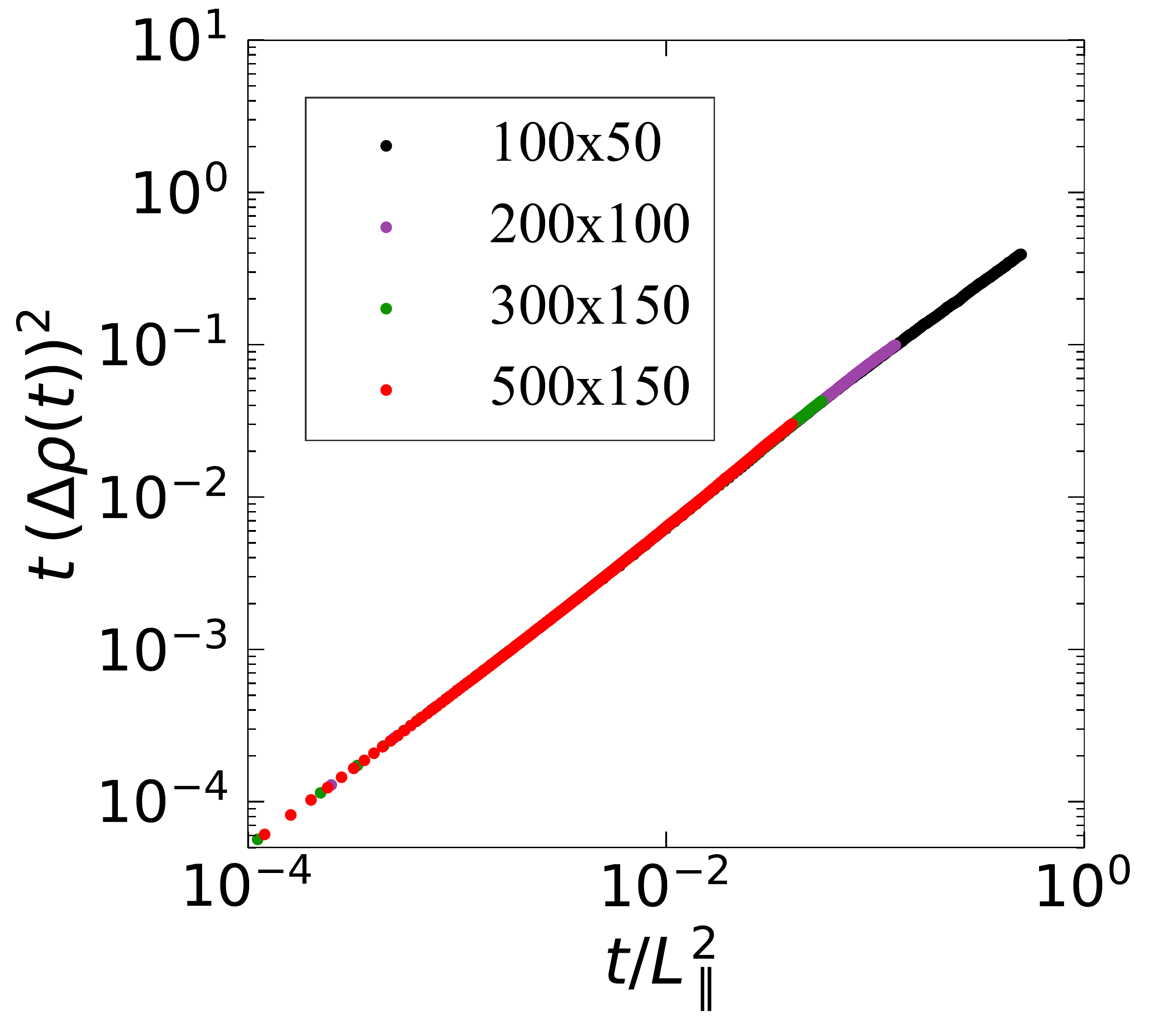}} \hfill
\subfloat[\label{fig:denfluct_critKLS}]{\includegraphics[width=0.48\columnwidth, 
	trim={0 0 0 0},clip]{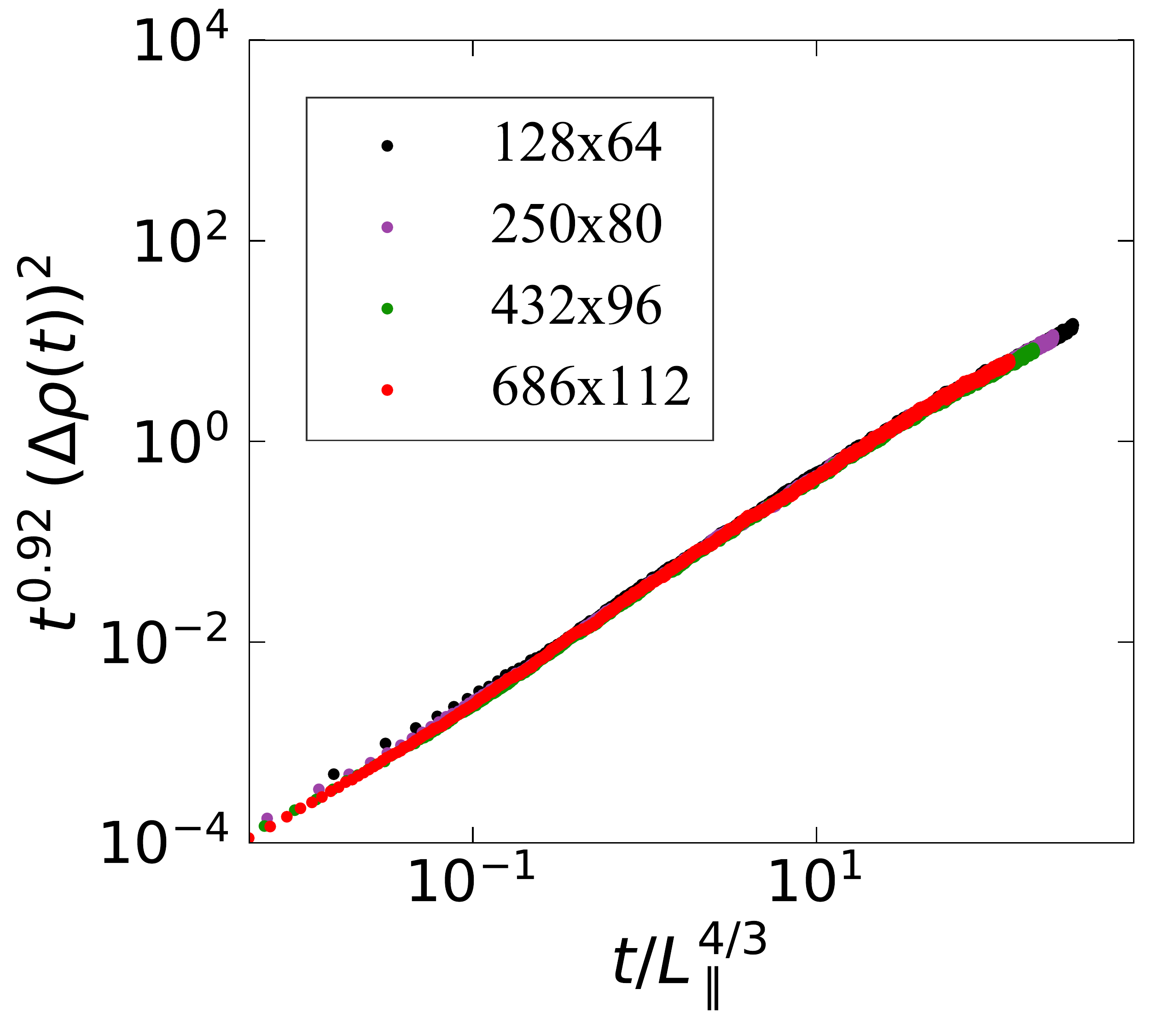}}
\caption{\label{fig:denfluct} 
	Finite-size scaling for the particle density fluctuations in the critical subsystem of the 
	two-temperature KLS driven lattice gas with parallel temperature interfaces. 
	The subsystem temperatures are: (a) $T_{\rm hot} = 5.0$, $T_{\rm cold} = 2.0$; 
	(b) $T_{\rm hot} = 5.0$, $T_{\rm cold} = T_c$, with $T_c^{128x32} = 0.782$, 
	$T_c^{250x40} = 0.788$, and $T_c^{432x48} = 0.794$.
	The data were averaged over $2,500$ independent realizations.}
\end{figure*}
As is demonstrated in Fig.~\ref{fig:denfluct}, the particle density fluctuations in the critical 
subsystem follow the scaling form
\begin{equation}\label{eqn:den_scaling}
     \left( \Delta \rho(t, L_{\parallel}, L_\perp) \right)^2 \sim t^{-b} \hat\varrho
     \left( t / L_{\parallel}^{z_{\parallel}}, L_{\parallel} / L_{\perp}^{1 + \Delta} \right) ,
\end{equation}
where $z$ is the dynamical critical exponent and $b$ denotes another scaling exponent, 
which stems from the density correlations.
Finite-size scaling renders Eq.~\eref{eqn:KLSscalingform} into \cite{Tauber_2014}
\begin{equation}\label{eqn:KLSscalingfinitesize}
     C\left(\tau, x_\parallel, x_\perp, t, L_{\parallel}, L_{\perp} \right) = 
     t^{-\zeta} \ {\hat C}\left( \tau |L_\perp|^{1 / \nu}, 
     \frac{x_\parallel}{L_{\parallel}}, \frac{x_{\perp}}{L_{\perp}}, 
     \frac{L_{\parallel}}{L_{\perp}^{1 + \Delta}}, 
     \frac{t}{L_\parallel^{z_\parallel}} \right) ,
\end{equation}
with $\zeta = (d + \Delta - 2 + \eta) / z = \Delta / z$ in $d = 2$ dimensions. 
The normalized particle density fluctuations \eref{eqn:denfluct} follow directly from the 
spatial integral of the correlation function; hence we expect in $d$ dimensions
\begin{align} \label{eqn:final_den_scaling}
     \left( \Delta \rho(t, L_{\parallel}) \right)^2 &\sim \int 
     C\left(\tau, x_\parallel, x_\perp, t, L_{\parallel}, L_{\perp} \right) dx_\parallel 
     d^{d-1}x_\perp / (L_\perp^{d-1} L_\parallel)^2 \\
     & \sim t^{- (2 d + 2 \Delta - 2 + \eta)/z}
     \hat\varrho(t/L_{\parallel}^{z_\parallel},L_{\parallel}/L_{\perp}^{1+\Delta}) \ , 
     \nonumber 
\end{align}
and thus obtain $b = 2 (d - 1 + \Delta + \eta / 2) / z = 2 (1 + \Delta) / z$ for $d = 2$.

Remarkably, the scaling exponents in Eq.~\eref{eqn:den_scaling} appear to change when 
the temperatures of the subsystems are varied. 
As shown in Fig.~\ref{fig:denfluct_TASEP}, when both temperatures are set well above 
$T_c$, the temporal evolution of the fluctuations satisfies finite-size scaling with exponents 
$z = 2$ and $b = 1$. 
Moreover, the dynamical scaling of the particle density fluctuations in this case does not 
depend on the choice of the system size aspect ratio $L_{\parallel} / L_{\perp}$, suggesting 
the absense of the manifest anisotropic scaling, i.e., $\Delta = 0$ for 
$T_{\rm hot}, T_{\rm cold} \gg T_c$. 
Indeed, these are just the (T)ASEP scaling exponents in two dimensions (neglecting 
logarithmic corrections at its upper critical dimension $d_c = 2$), as listed in 
Table.~\ref{tab:table}. 
This exponent match with the finite-size scaling in our data demonstrates that both 
subsystems  of the two-temperature KLS model with parallel temperature interfaces follow 
the (T)ASEP dynamical scaling behavior when the temperatures $T_{\rm hot}$ and 
$T_{\rm cold}$ in either region are chosen much higher than the KLS critical temperature 
$T_c$, and consequently the attractive ferromagnetic Ising interactions largely irrelevant.

In contrast, when we choose the subsystems' temperatures as $T_{\rm hot} \gg T_c$, but
$T_{\rm cold} = T_c$, the particle density fluctuations display very different finite-size 
scaling behavior. 
Indeed, the scaling collapse becomes sensitive to the system size aspect ratio, signaling the 
presence of strong anisotropic scaling. 
Once we select the appropriate size aspect ratio 
$L_{\parallel} / (a L_{\perp})^{1+\Delta} = 1/256$ for the critical subsystem with the KLS 
critical anisotropy exponent $\Delta = 2$ in two dimensions, we observe convincing data 
scaling collapse for the particle density fluctuations, as evidenced in 
Fig.~\ref{fig:denfluct_critKLS} with exponents $z_\parallel = z / (1+\Delta) = 4/3$ and 
$b \approx 0.92$. 
While $z = 4$ is indeed equal to the KLS dynamical critical exponent, the collapse exponent 
deviates from the value $b = 3/2$ predicted by our scaling ansatz. 
We attribute this discrepancy to strong finite-size corrections characteristic of the critical KLS
model, along with the interference of the distinct boundary dynamics, which is prominent for 
$T_{\rm hot} \gg T_c$ and $T_{\rm cold} = T_c$.
(We note that the mean-field value for the KLS model, with upper critical dimension 
$d_c = 5$ and $\Delta = 1$, is $b = 5 / 2$.)
Still, the at least approximate scaling of the particle density fluctuations with the KLS critical 
exponents indicates the prominence of critical fluctuations in the two-temperature KLS 
system for the above range of subsystem temperatures, and that in fact these long-range
correlations control the hybrid KLS dynamics at least at the system sizes and time scales our
simulations could access. 
We believe that the critical KLS scaling becomes established in this situation because the 
transverse particle current is prominent at the subsystem interfaces for those temperature 
values, and particles are being screened from the temperature boundaries. 
This leads to an effective (albeit incomplete) decoupling of the hotter and colder regions, 
allowing the strong KLS fluctuations to build up in the critical subsystem. 

\subsection{\label{sec:sublevel33} Entropy production rate}

\begin{figure*}[t!]
%\captionsetup[subfloat]{slc=off,margin={1cm,0cm}}, valign=c
\centering
\subfloat[\label{fig:entropyprod_hot}]{\includegraphics[width=0.5\columnwidth, 
	trim={0 0 0 0},clip]{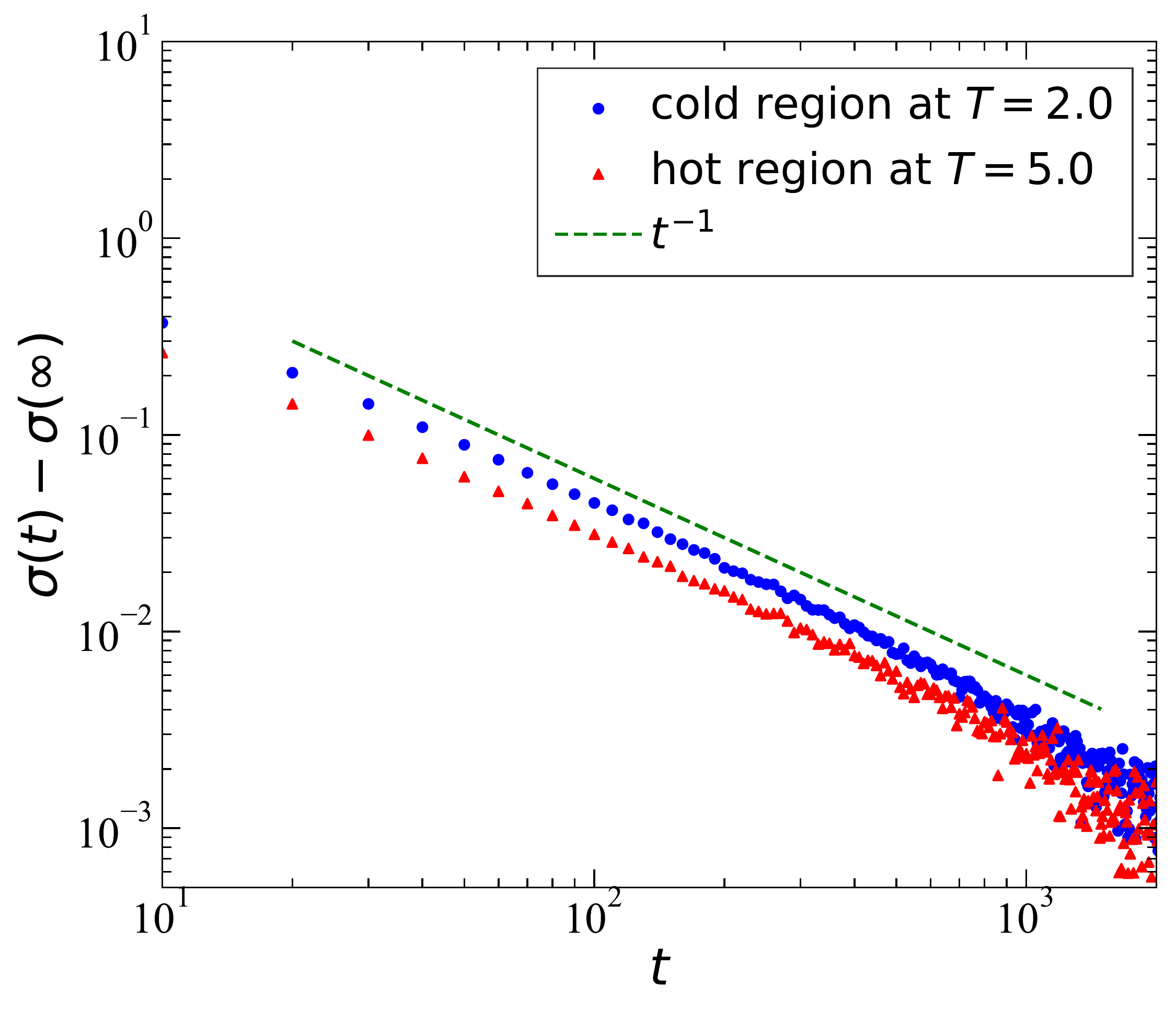}} \hfill
\subfloat[\label{fig:entropyprod_Tc}]{\includegraphics[width=0.5\columnwidth,
	trim={0 0 0 0},clip]{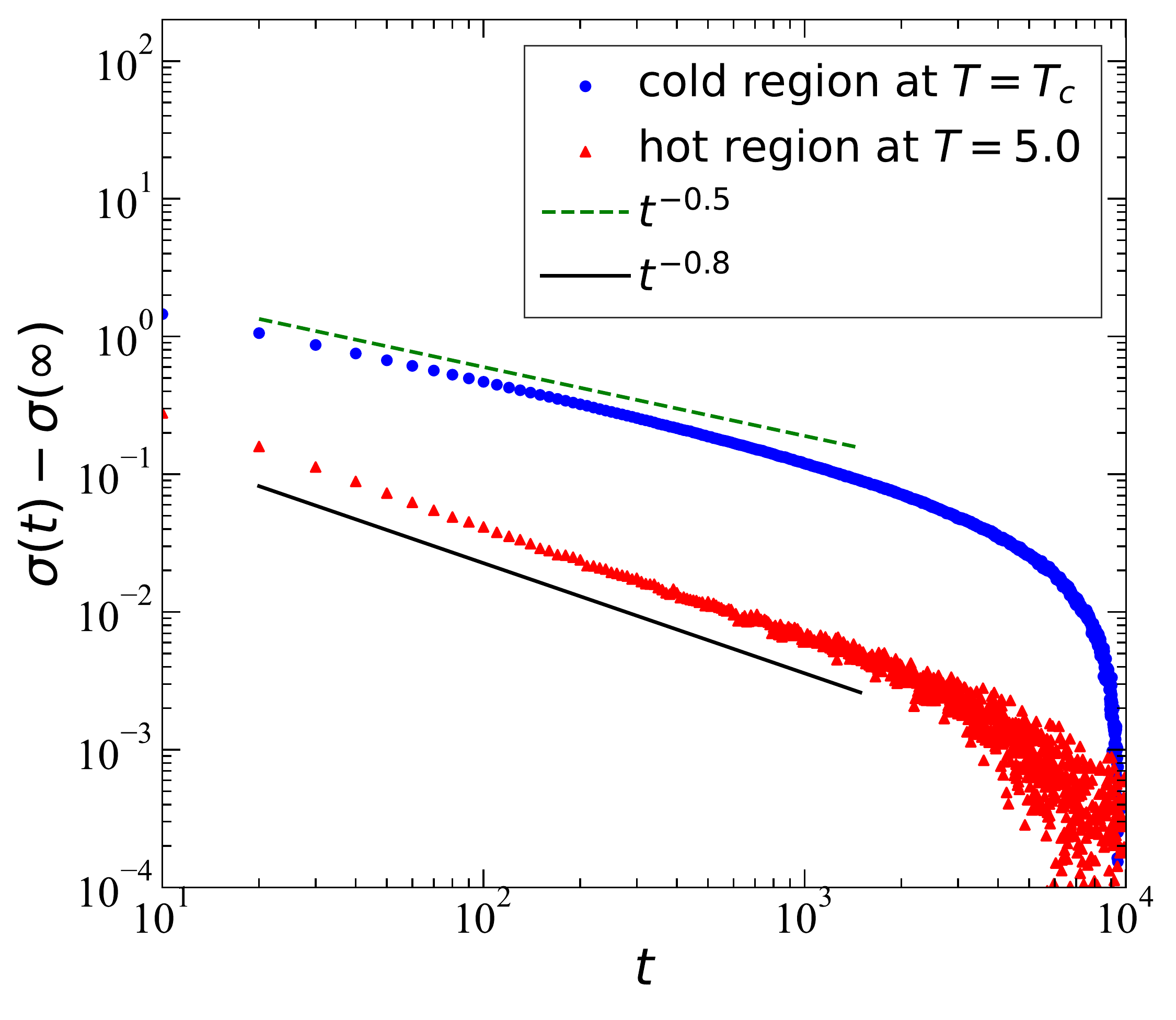}}
	\vskip\baselineskip
	\vspace{-0.5cm}
\subfloat[\label{fig:EPR_KLS_TASEP}]{\includegraphics[width=0.49\columnwidth, 
	trim={0 0 0 0},clip]{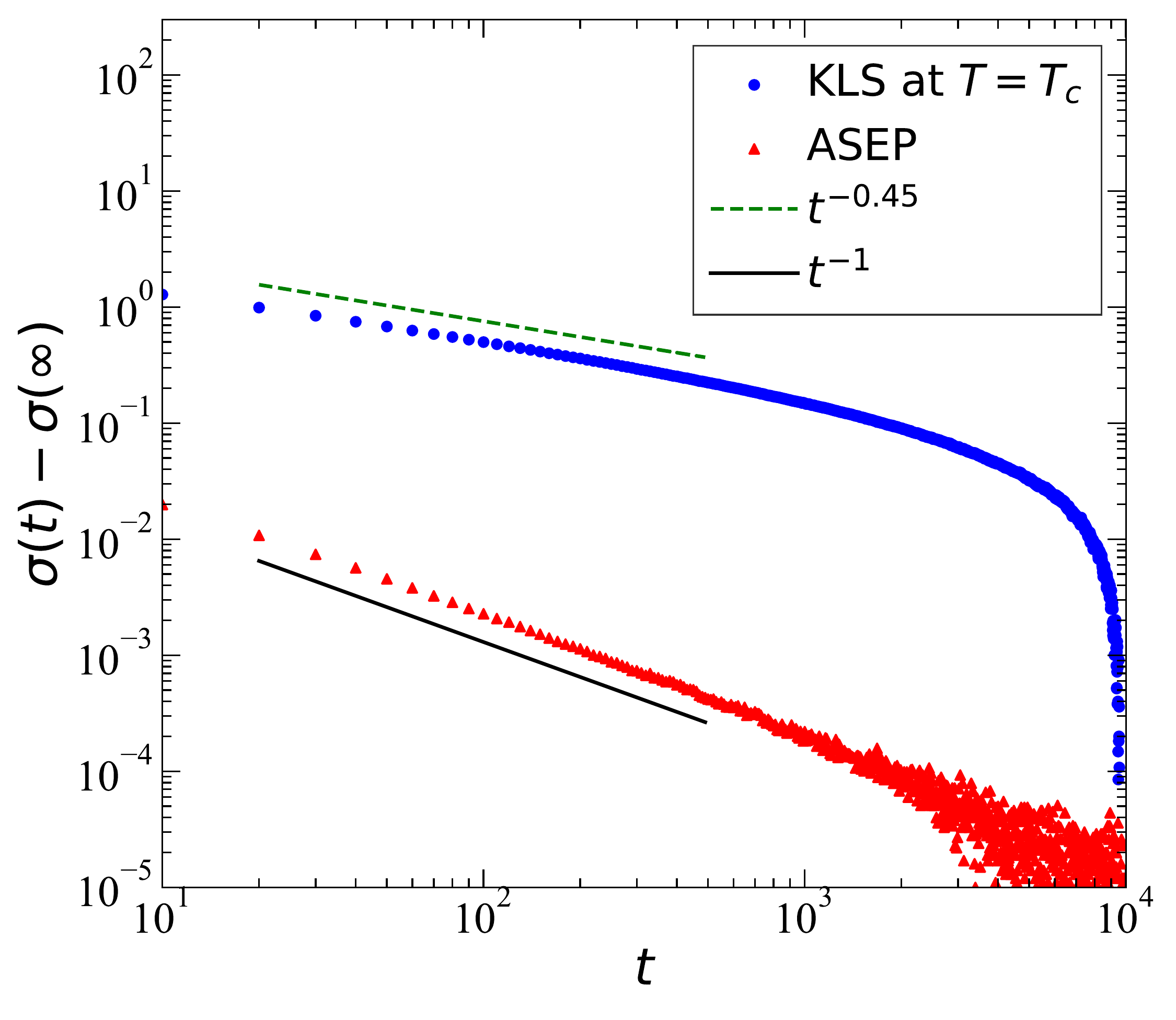}} \hfill
\subfloat[\label{fig:Current_KLS_TASEP}]{\includegraphics[width=0.5\columnwidth,
	trim={0 0 0 0},clip]{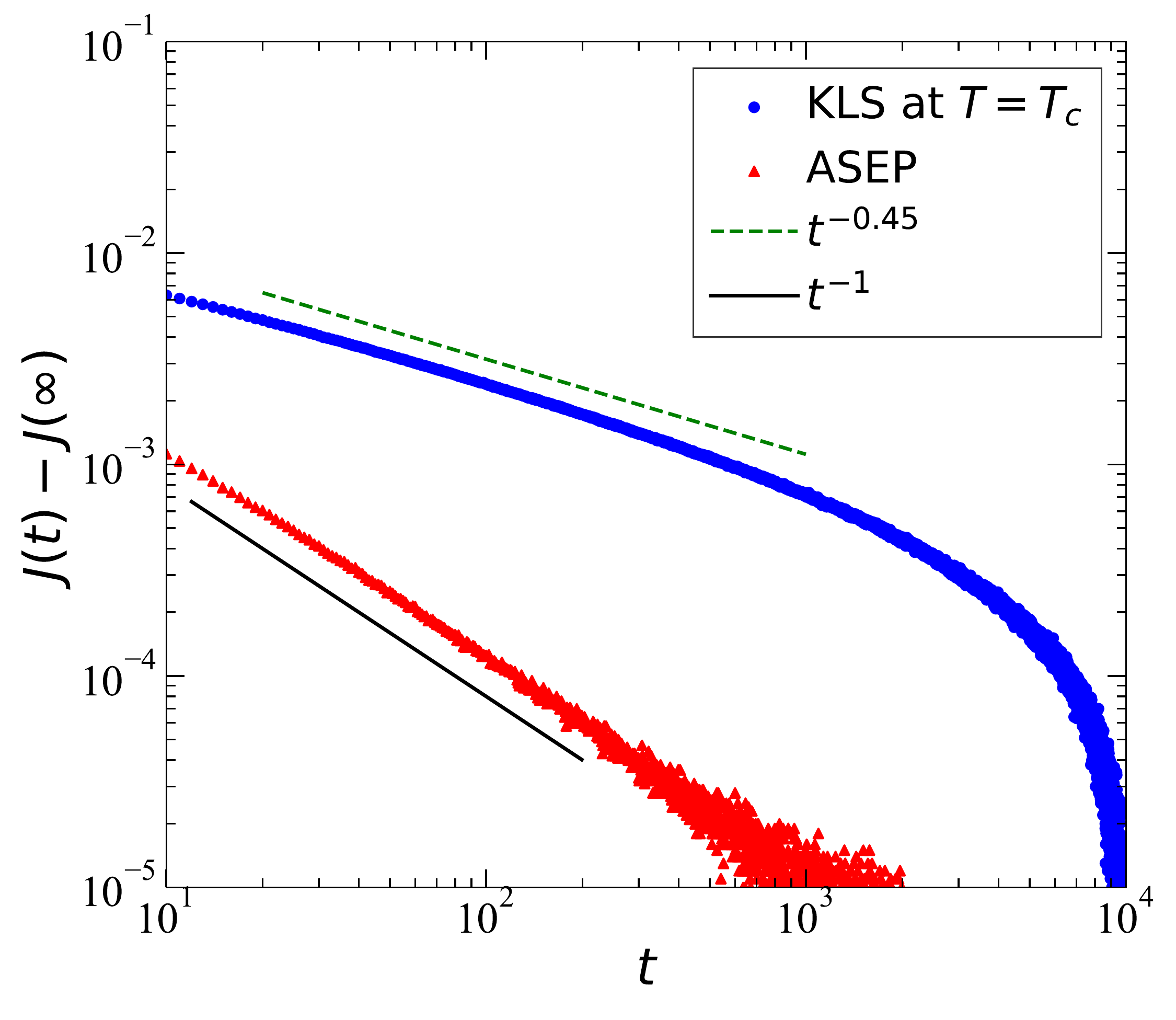}}
\caption{\label{fig:entrprod} 
	(a) The entropy production rate per volume in the two-temperature KLS driven lattice 
	gas with parallel temperature interface in both subsystems for $T_{\rm hot} =5.0$ and 
	$T_{\rm cold}=2.0$ after the saturation value $\sigma(\infty)$ has been subtracted. 
	(b) The entropy production rate per volume in both temperature regions for 
      $T_{\rm hot} = 5.0$ and $T_{\rm cold} = T_c^{250x40}=0.788$ after the saturation 
	value $\sigma(\infty)$ has been subtracted.
	(c) The entropy production rate per volume and (d) net particle current in the standard 
	critical KLS model at $T=T_c^{250x40}=0.788$ and in the ASEP after the saturation 
	values $\sigma(\infty)$ and $J(\infty)$ have been subtracted.
	The system dimensions are: $L_{\parallel} = 250$, $L_{\perp}=80$ with $1:1$
	subsystem ratio for both (a) and (b); $L_{\parallel} = 250$, $L_{\perp} = 40$ for the 
	standard KLS model, and $L_{\parallel} = L_\perp = 250$ for the ASEP in (c), (d).
	The drive strength is chosen to be $E=100$. 
	The ASEP hopping rate along the drive is set to $p_\parallel =0.999$, and in the
	opposite direction $q_\parallel = 1- p_\parallel =0.001$. 
	All data were averaged over $100,000$ independent realizations.}
\end{figure*}
To reinforce our argument that critical fluctuations are observable in the hybrid 
two-temperature KLS system when the subsystem temperatures are set to 
$T_{\rm hot} \gg T_c$ and $T_{\rm cold} = T_c$, we have measured the entropy 
production rate in both subsystems. 
The non-negative entropy production rate per volume is defined on the two-dimensional 
lattice 
as \cite{Jiang:2004}
\clearpage
\begin{align}\label{eqn:EPR}
     \sigma(t) = \frac{1}{2 L_\parallel L_\perp} \sum_{\mathcal{C}, \mathcal{C}'}
     \sum_{i,j}^{L_{\parallel}, L_{\perp}} & \left[ P_{ij} (\mathcal{C},t) 
     W_{ij}(\mathcal{C} \to \mathcal{C}', t) - P_{ij}(\mathcal{C}',t) 
     W_{ij}(\mathcal{C}' \to \mathcal{C}, t) \right] \nonumber \\
     &\times \ln \frac{P_{ij}(\mathcal{C},t) W_{ij}(\mathcal{C} \to \mathcal{C}', t)}
    {P_{ij}(\mathcal{C}',t) W_{ij}(\mathcal{C}' \to \mathcal{C}, t)} \ge 0 \ ,
\end{align}
where the sum extends over all $2^6$ possible nearest-neighbor configurations of any
selected site pair chosen for particle-hole exchange.
We compute this quantity in our simulations by obtaining the probability 
$P_{ij}(\mathcal{C},t)$ of the each configuration to occur at each lattice site at time $t$. 
The logarithm factor in Eq.~\eref{eqn:EPR} prohibits us from computing the entropy 
production rate with infinite drive strength, as it sets the hopping rate against the drive to 
zero. 
Thus, to obtain the entropy production rate we set the drive strength to a very large, but
finite value $E = 100$.

Measuring the entropy production rates for the hotter and cooler subsystems separately, we 
found that they quickly decay to steady-state values after we quench the system from random 
initial conditions or infinite temperature to $T_{\rm hot}, T_{\rm cold}$, subsequently 
holding the temperatures constant at $T_{\rm hot}, T_{\rm cold} \gg T_c$. 
Since the density correlations in this (T)ASEP temperature regime are very short-ranged, one 
would indeed expect this fast relaxation from a generic initial configuration to a low-level, 
external noise-driven stationary entropy production rate. 
However, if we initiate our simulations with highly correlated ``checkerboard'' configurations
with alternating filled and empty sites, a state that is quite distinct from any generic TASEP
configuration \cite{Daquila:2011}, then the entropy production rate decays according to a 
power law $\sim t^{-1}$ in both subsystems for $T_{\rm hot}, T_{\rm cold} \gg T_c$, as 
shown in Fig.~\ref{fig:entropyprod_hot}.
Yet when the temperature in the colder region is maintained exactly at the critical value 
$T_{\rm cold} = T_c$, we observe a slower algebraic decay of the entropy production rate 
with decay exponent $\approx 0.5$ in the critical subsystem, and decay exponent 
$\approx 0.8$ in the hotter subsystem, as depicted in Fig.~\ref{fig:entropyprod_Tc}, 
followed by a finite-size exponential cut-off.

To address the question whether the power-law decay of the entropy production rate in the
critical subsystem is dominated by the KLS critical fluctuations, and to better grasp how the 
hotter subsystem's dynamics affects the non-equilibrium relaxation of the critical subsystem 
to its steady state, we measured the entropy production in the standard, homogeneous 
two-dimensional KLS and ASEP models, with biased hopping rates along and against the
drive set to $p_{\parallel}=0.999$ and $q_{\parallel} = 1-p_\parallel = 0.001$, respectively. 
To access the nontrivial relaxational regime in the ASEP, we prepared the system agin in a 
unique correlated initial state, placing the particles with alternating occupation on the lattice 
in a checkerboard fashion \cite{Daquila:2011}. 
As shown in Fig.~\ref{fig:EPR_KLS_TASEP}, the power law for the entropy production rate 
in the critical subsystem follows very closely the analagous algebraic decay for the standard 
KLS model at criticality, indicating that critical fluctuations persist in the KLS system, even 
when it is being coupled to a region with hotter temperature $T_{\rm hot} > T_c$. 
Similarly, if both subsystem temperatures are set well above $T_c$, the entropy production 
rate decays with the ASEP exponent, as one would expect in the KLS high-temperature phase. 

From our simulations it is evident that these nontrivial power-law decay exponents must fall 
into the respective critical KLS and TASEP universality classes. 
The expression for the entropy production rate in Eq.~\eref{eqn:EPR} clearly suggests that 
its temporal evolution should be closely tied to the non-equilibrium relaxation of the net 
particle current following the quench. 
In the absence of the external drive, the entropy production stems from the change in the 
system's energy, due to its coupling with its surroundings \cite{Spohn:1991, Cates:2020}. 
Yet for the driven lattice gases under consideration here, $\sigma(t)$ is dominated by the 
change of the entropy within the system, which in turn equals the drive strength multiplied 
with the mean current in the system \cite{Spohn:1991}. 
This connection between the entropy production rate and the average current is confirmed 
nicely in our simulation data through the match of the corresponding decay exponents. 
We show in Fig.~\ref{fig:Current_KLS_TASEP} that as the mean particle current and the 
entropy production rate approach their steady state values, the power laws for the average 
particle current difference $j(t) - j(\infty)$ and for the entropy production rate difference 
$\sigma(t) - \sigma(\infty)$ are indeed characterized by the same decay exponents, for both 
the ASEP and KLS models. 

Furthermore, the origin of these decay exponents becomes apparent once we relate the 
entropy production rate with the nearest-neighbor equal-time correlation function. 
One can see how these two quantities are connected after expanding the exponential factor 
in the Kawasaki exchange rates in Eq.~\eref{eqn:KLSrates} for $E = 0$: 
From the Ising Hamiltonian \eref{eqn:IsingHamilt} it is clear that the transition rate 
difference 
$W_{ij}(\mathcal{C} \to \mathcal{C}', t) - W_{ij}(\mathcal{C}' \to \mathcal{C}, t)$ will 
thus become to leading order proportional to the nearest-neighbor equal-time correlation 
function.
This makes sense, since the entropy production rate originates from the temporal changes in 
the system's energy, which is here characterized by the nearest-neighbor equal-time 
correlation function in the thermodynamic limit. 
Indeed, a similar connection between the entropy production rate and the rate of free energy 
change has been established previously for non-driven Ising spin systems \cite{Cates:2020}. 
Consequently, we posit that the ASEP and KLS entropy production power laws observed in 
Fig.~\ref{fig:EPR_KLS_TASEP} are determined by the temporal scaling of the 
nearest-neighbor equal-time correlation function, i.e., 
$\sigma(\infty) - \sigma(t) \sim t^{-\zeta}$, where $\zeta$ is just the decay exponent from 
Eq.~\eref{eqn:KLSscalingfinitesize}.

% =====================================================
% IV. Summary and Conclusion
% ===================================================== 
\section{\label{sec:level4} Conclusion}

In this work we have numerically studied a two-temperature KLS driven lattice gas, with the 
temperature interfaces aligned along the external drive direction.
We have demonstrated that it is possible to control the scale-invariant dynamics in the 
system by changing the temperature(s) locally, i.e., by adjusting the strength of the vortex 
sheets at the temperature boundaries.  
In contrast to our previous work, where splitting the lattice into two temperature regions 
with transverse temperature boundaries induced a density phase separation in the hotter
subsystem, here we observe that orienting the interfaces parallel to the drive and net 
current maintains homogeneity in each subsystem and preserves its scale-invariant 
dynamics.

Even though the formation of the boundary-induced vortex sheets terminates the
subsystems' independent initial aging kinetics, their coupling through particle exchange
across the interfaces allows us to observe interesting finite-size scaling for the resulting 
particle density fluctuations. 
We have shown that for two distinct temperature regimes, namely for 
$T_{\rm hot}, T_{\rm cold} \gg T_c$ and $T_{\rm hot} \gg T_c, T_{\rm cold} =T_c$,
these particle density fluctuations scale with the (T)ASEP and critical KLS dynamical scaling 
exponents, respectively. 
Moreover, in the latter case we have observed the power-law decay exponent of the entropy 
production rate in the critical subsystem to be very close to the corresponding value 
measured for the standard critical KLS model, which in turn is asymptotically governed by
the scaling of the density autocorrelation function.
Although we have gathered substantial evidence that the dynamics of the critical subsystem 
is governed by the KLS critical exponents, we cannot completely exclude the possibility that
 for the latter temperature choice, the system might be arrested in an exceedingly slow 
crossover regime, that critical fluctuations will eventually terminate, and the system 
ultimately display TASEP scaling behavior. 
We plan to explore the possibility of controlling the critical KLS scaling behavior in various 
two-temperature hybrid KLS models in greater detail in future studies. 
In particular, we are interested in varying the subsystems' lateral aspect ratio, perpendicular
to the external drive, and intend to investigate to what extent one could shrink either of the 
temperature regions while preserving their scale-invariant dynamics.

After considering both the transverse and the parallel alignment variants for the interfaces
in the two-temperature KLS driven lattice gas, it is natural to ask which of these systems'
distinct physics will prevail in an intermediate case, when the temperature boundaries are 
tilted relative to the external drive, but not orthogonal to it. 
For such ``diagonal'' temperature interfaces, we expect to observe similar density phase 
separation as for the case of fully transverse subsystem boundaries; yet, in addition, we 
would also anticipate to see vortex current sheets located right at the temperature interfaces. 
A detailed study is required to answer questions such as whether the phase separation 
interface would be localized at the center of the hot subsystem, and whether this generalized 
diagonal two-temperature KLS variant still possesses dynamical scaling properties when the 
cooler subsystem is held at the critical temperature. 

% =====================================================
% Appendix: BROKEN SYMMETRY
% ===================================================== 
\section*{Appendix: Interface hopping rates with broken particle-hole symmetry}

To satisfy our curiosity, we have considered another (simpler) choice for the hopping rates 
across the temperature interfaces that explicitly violates particle-hole symmetry, namely:
\begin{equation} \label{eqn:across_broken}
     W(\mathcal{C} \to \mathcal{C}'; T_1 \to T_2) \propto \exp
     \left[- \left( \frac{H(\mathcal{C}')}{T_2} - \frac{H(\mathcal{C})}{T_1} \right) \right] ,
\end{equation}
where $H(\mathcal{C})$ and $H(\mathcal{C}')$ are the energy functions from 
Eq.~\eref{eqn:IsingHamilt}, $T_1$ is the temperature of the subsystem that a 
\textit{particle} tries to leave, and $T_2$ is the temperature of the subsystem that this
particle attempts to enter.
\begin{figure*}
\centering
\includegraphics[width=0.65\columnwidth]{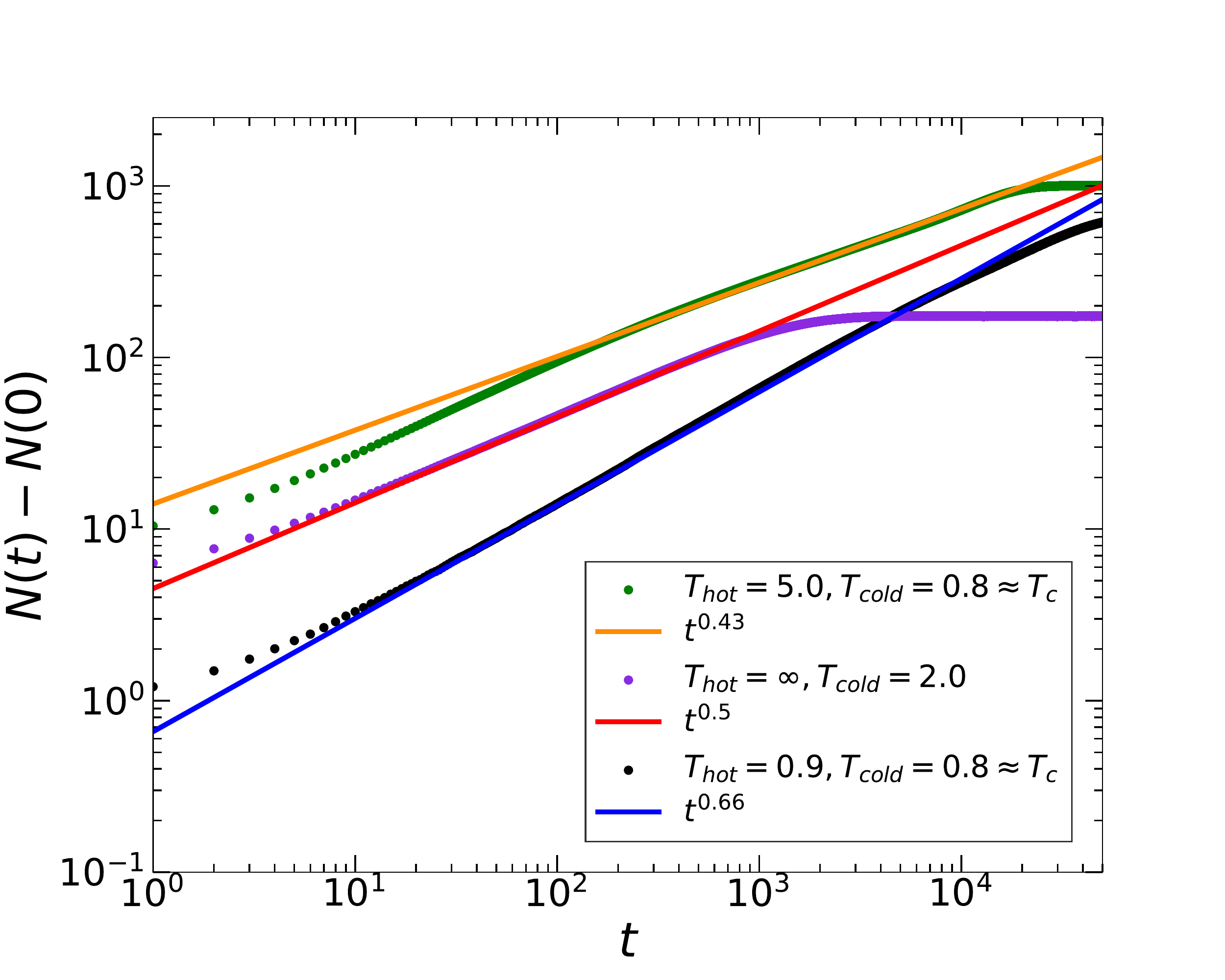}
	\caption{\label{fig:N_growth}
  	Net particle growth in the hot subsystem with time for different subsystem
      temperatures in the two-temperature KLS model with hopping rates across the 
      temperature boundaries that manifestly break particle-hole (Ising $Z_2$) symmetry. 
      The system size is $L_\parallel =  128$, $L_\perp = 64$, with $1:1$ subsystem size 
      ratio. 
      The data were averaged over $10,000$ independent realizations.}    
\end{figure*}

In the rate prescription (\ref{eqn:across_broken}), the temperature values $T_1$ and 
$T_2$ depend on the initial and the final positions of particle that tries to cross the 
temperature boundary. 
This choice manifestly breaks particle-hole symmetry and, as a result, induces a net flow of 
particles across the temperature interfaces into the hotter region. 
This counterintuitive outcome simply stems from the form of the hopping rates across the 
temperature boundaries (\ref{eqn:across_broken}), which makes it easier for particles to 
exit the critical subsystem rather than to enter it, leading to a transport bias into the hotter
region.
As shown in Fig.~\ref{fig:N_growth}, the particle excess in the hotter subsystem exhibits an 
intriguing power law growth with an exponent that depends on both regions' temperatures:
We observe slower growth with a smaller exponent value for $T_{\rm cold} = T_c$ and 
$T_{\rm hot} \gg T_c$, indicating that the particle exchange between the subsystems is 
hindered for these temepratures.
When both temperatures are set well above $T_c$, the motion in the transverse direction is 
purely diffusive and hence the power-law growth exponent is $1/2$, directly related to the 
transverse mean-square particle displacement. 
We cannot offer insights why we measure a $t^{2/3}$ algebraic growth if both 
$T_{\rm hot}, T_{\rm cold} \gg T_c$. 
We had initially surmised that this growth exponent might reflect a universal (T)ASEP 
feature, but had to discard that notion after observing distinct growth exponents for other
$Z_2$  symmetry-breaking forms of the interface hopping rates.
In fact, we have found that the growth exponent can take any value in the interval 
$[0.4:0.83]$, depending on the definition of the hopping rates across the temperature 
boundaries and both regions' temperatures. 
\begin{figure*}[t!]
%\captionsetup[subfloat]{slc=off,margin={1cm,0cm}}, valign=c
\centering
\subfloat[\label{fig:denprof_broken}]{\includegraphics[width=0.49\columnwidth, 
	trim={0 0 0 0},clip]{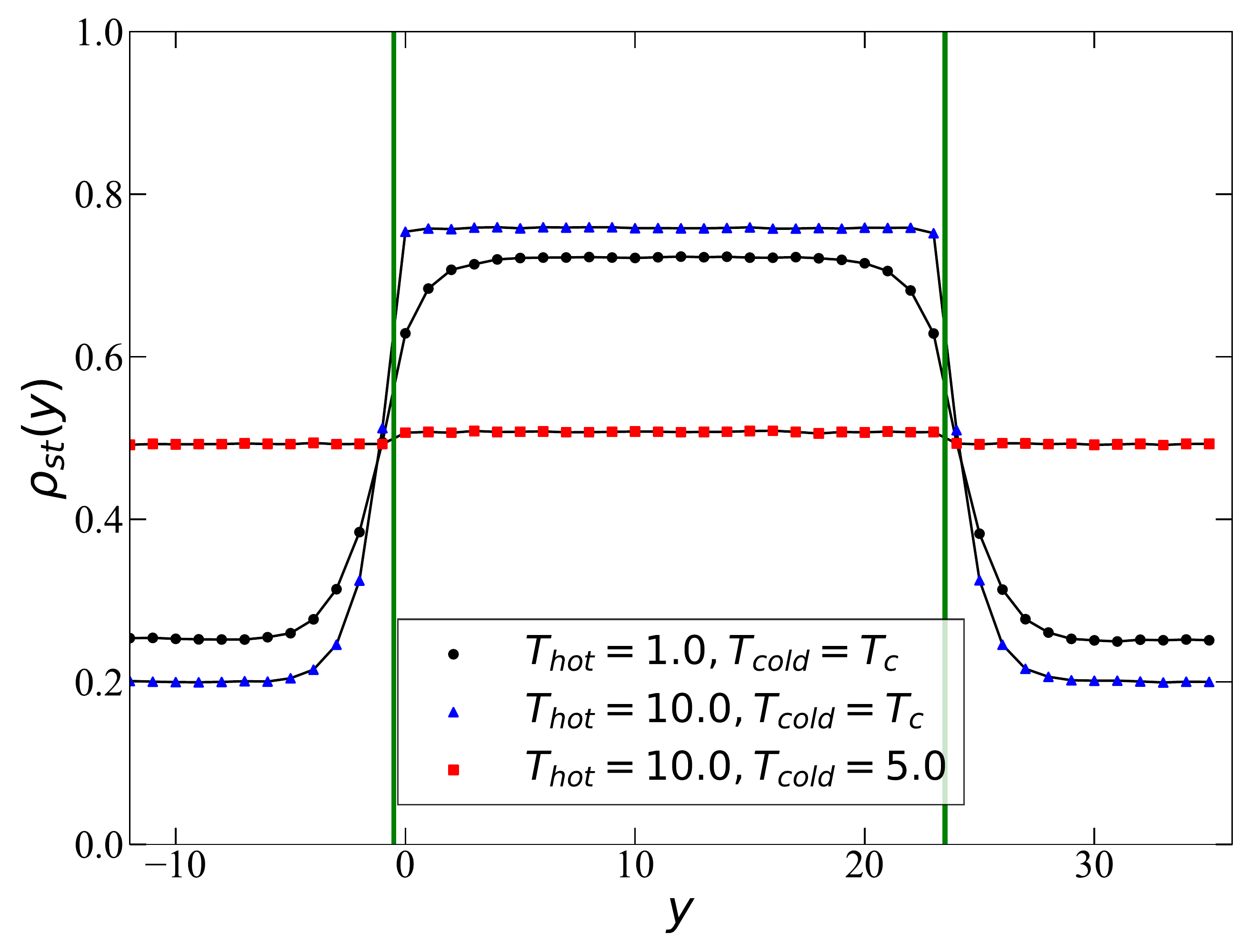}} \hfill
\subfloat[\label{fig:curprof_broken}]{\includegraphics[width=0.48\columnwidth,
	 trim={0 0 0 0},clip]{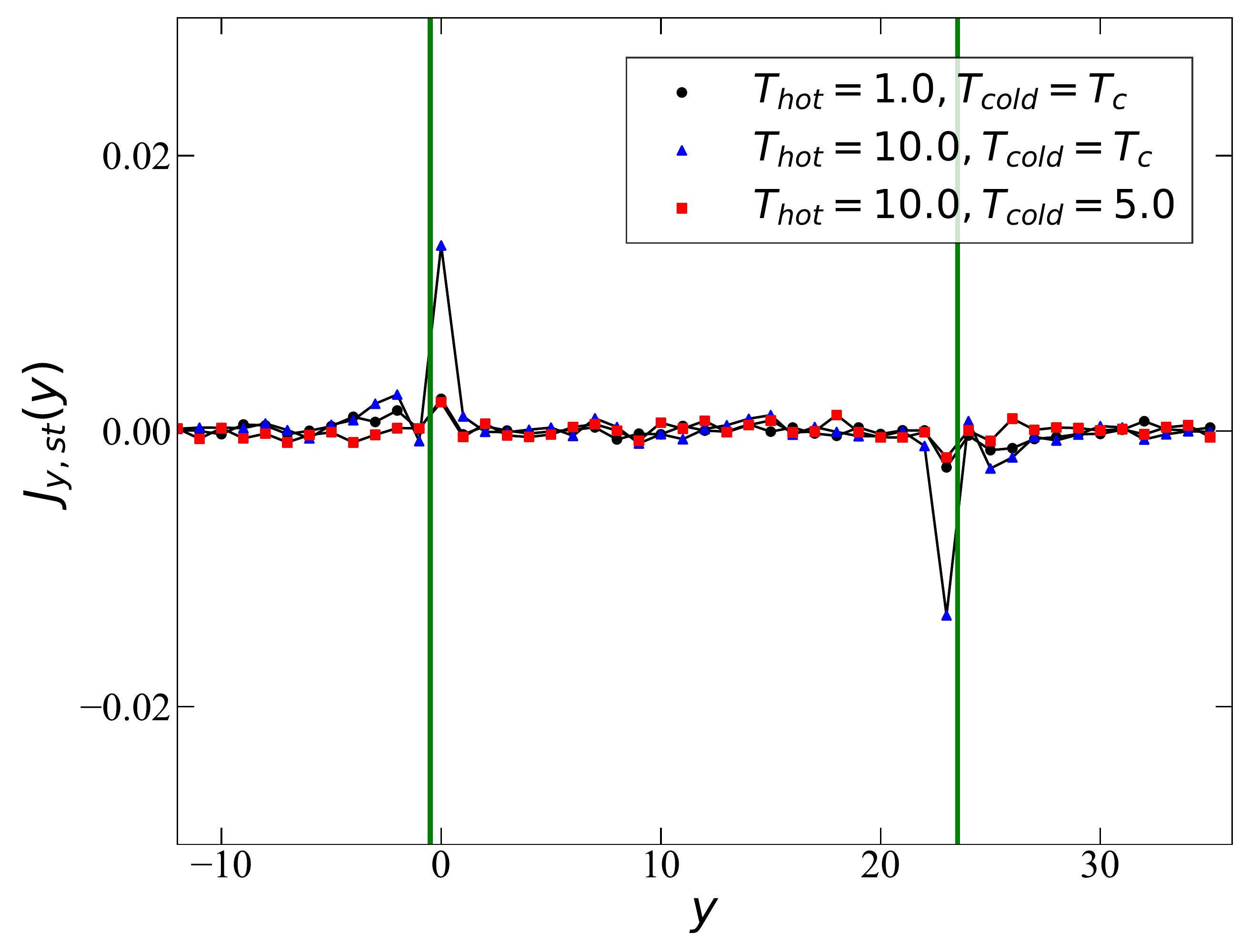}}
\caption{\label{fig:prof_broken} 
	(a) The density and (b) current profiles of the two-temperature KLS driven lattice gas 
	with parallel temperature interfaces for different subsystem temperatures 
	$T_{\rm hot}$ and $T_{\rm cold}$. 
	The lattice dimensions are $L_\parallel = 54$, $L_\perp = 48$, with subsystem size
	ratio $1:1$; the temperature interface boundaries are indicated by the green lines. 
	The hotter temperature subsystem is located in the center on these plots. 
	The data were averaged over $10,000$ independent realizations.}
\end{figure*}

Once the particle number growth in the hotter subsystem shown in Fig.~\ref{fig:N_growth} 
ceases, we presume the two-temperature KLS driven lattice gas to have reached its 
non-equilibrium steady state. 
The resulting steady-state density profile demonstrates that the hopping rates across the 
temperature boundaries that break particle-hole symmetry trigger a phase separation in the
particle density; the initially homogeneous system splits into a high-density region in the hot 
subsystem, and a low-density region in the critical subsystem, with sharp density kinks 
located at the temperature interfaces. 
As depicted in Fig.~\ref{fig:denprof_broken}, the height of the density kinks grows as the 
transverse current at the temperature boundaries increases (Fig.~\ref{fig:curprof_broken}). 
We remark that finite-size effects are likely significant in this clogged stationary state, where
mutual exclusion at the interfaces ultimately terminates the further dynamical evolution of the
system.
\begin{figure*}[ht]
%\captionsetup[subfloat]{slc=off,margin={1cm,0cm}}, valign=c
\centering
\subfloat[\label{fig:curvecplot_broken_1.0}]{\includegraphics[width=0.48\columnwidth, 
	trim={0 0 0 0},clip]{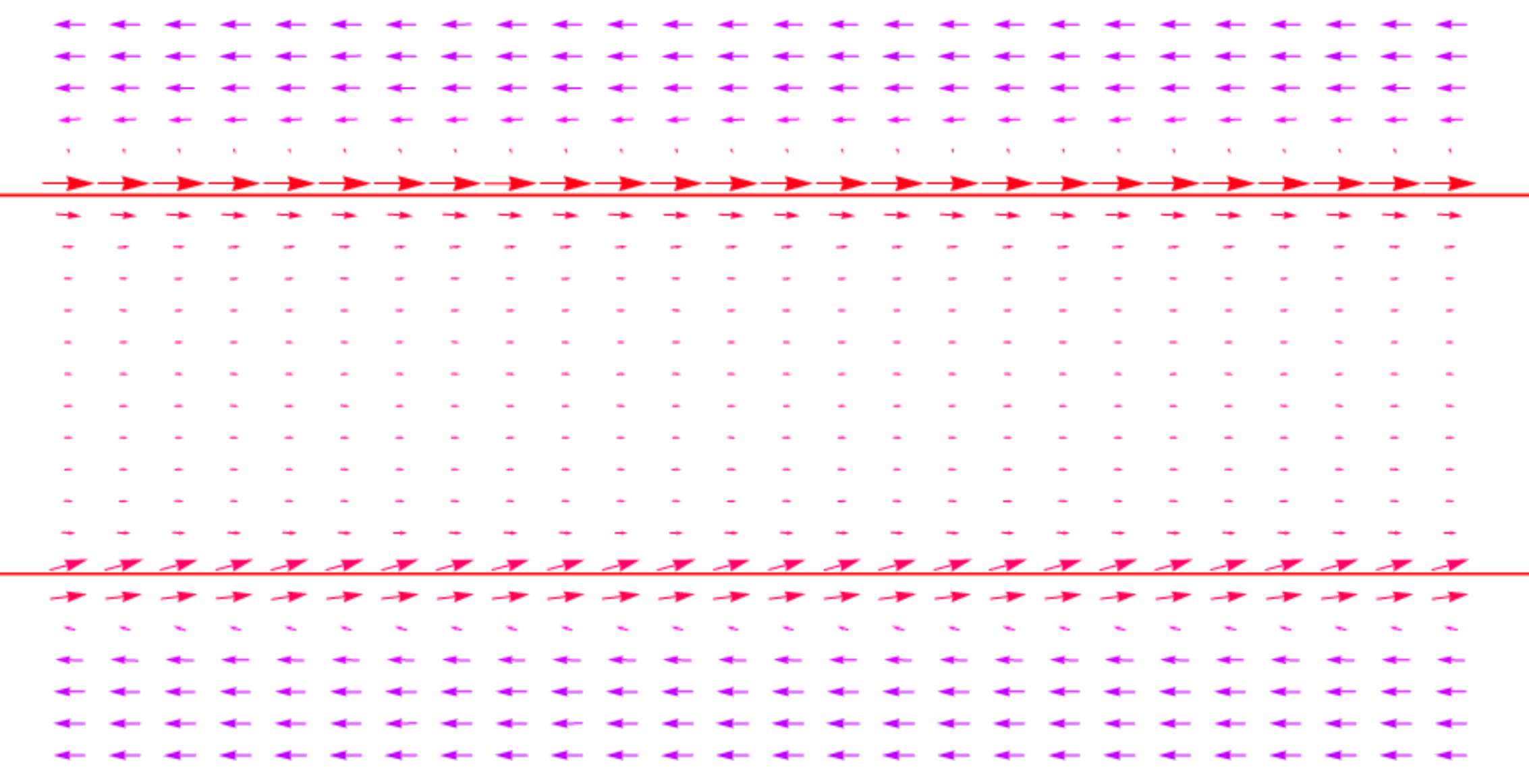}} \hfill
\subfloat[\label{fig:curvecplot_broken_10.0}]{\includegraphics[width=0.48\columnwidth,
	trim={0 0 0 0},clip]{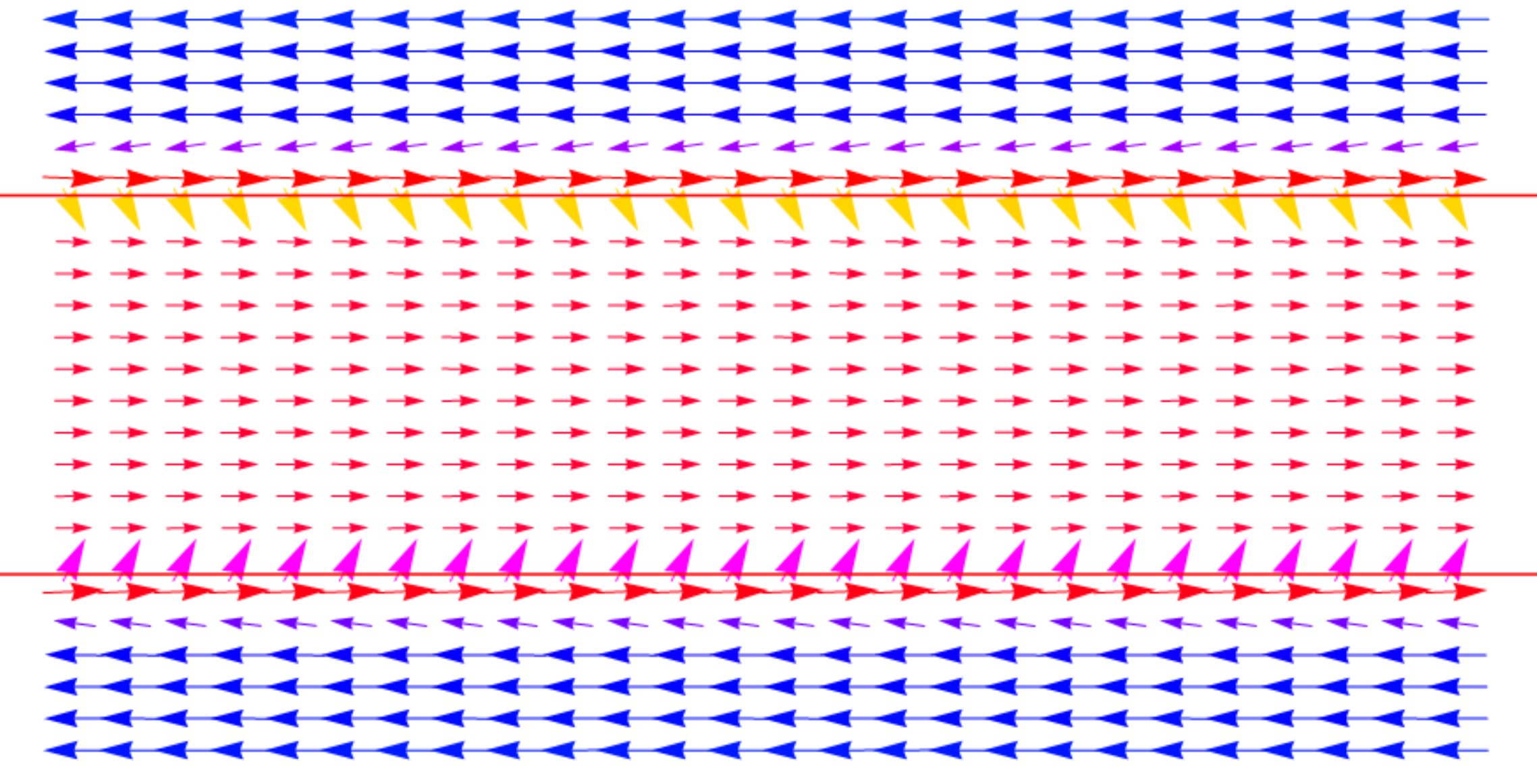}}
\caption{\label{fig:curvecplot_broken} 
	The coarse-grained current vector plots for the two-temperature KLS in the steady state 
	after the net current has been subtracted from each lattice point. 
	The subsystem temperatures are set at: (a) $T_{\rm hot} = 1.0$, 
	$T_{\rm cold} = T_c^{54\times 24} = 0.773$; 
	(b) $T_{\rm hot} = 5.0$, $T_{\rm cold} = T_c^{54\times 24} = 0.773$. 
	The system size is $L_\parallel =  54$, $L_\perp = 48$, with the hot temperature 
	subsystem located in the center of the plots; the red lines indicate the positions of the 
	temperature boundaries. 
	The data were averaged over $28,000$ Monte Carlo steps and over $10,000$ 
	independent realizations.}
\end{figure*}

Similarly to the particle-hole symmetric case, the density kink height increases with the 
difference between the mean bulk hopping rate from (\ref{eqn:KLSrates}) at $T_2$ and 
the average hopping rate across the interface given now by Eq.~\eref{eqn:across_broken}. 
However, in contrast to the symmetric rate choice, a second minor peak is visible in 
Fig.~\ref{fig:curprof_broken} right outside of the hotter region, signaling the presence of 
particle blockages at the interfaces.
A signature of ensuing particle clogging at the temperature boundaries can be also seen in 
the current vector plot in Fig.~\ref{fig:curvecplot_broken}, where near the interfaces, just 
outside of the hotter subsystem, there no significant transverse current is discernible, 
indicating limited particle influx into the hot region. 
As for particle-hole symmetric hopping rates, boundary effects become amplified when the 
difference between the mean hopping rates approaches its maximum value, i.e., when the 
subsystem temperatures are set to their extreme limits $T_{\rm hot} \to \infty$ and 
$T_{\rm cold} = T_c$.

\section*{Acknowledgments}
We would like to thank Michel Pleimling for fruitful discussions and valuable suggestions. 
Research was sponsored by the U.S. Army Research Office and was accomplished under 
Grant Number W911NF-17-1-0156. 
The views and conclusions contained in this document are those of the authors and should 
not be interpreted as representing the official policies, either expressed or implied, of the 
Army Research Office or the U.S. Government. 
The U.S. Government is authorized to reproduce and distribute reprints for Government 
purposes notwithstanding any copyright notation herein.

\section*{References}

\bibliographystyle{iopart-num-mod}
\bibliography{mybib}

\providecommand{\newblock}{}
\begin{thebibliography}{10}
\expandafter\ifx\csname url\endcsname\relax
  \def\url#1{{\tt #1}}\fi
\expandafter\ifx\csname urlprefix\endcsname\relax\def\urlprefix{URL }\fi
\providecommand{\eprint}[2][]{\url{#2}}
% Bibliography created with iopart-num v2.1
% /biblio/bibtex/contrib/iopart-num

\bibitem{Katz:1983}
Katz S, Lebowitz J~L and Spohn H, {\em Phase transitions in stationary
  nonequilibrium states of model lattice systems\/},  1983 {\em Phys. Rev. B\/}
  {\bf 28} 1655--1658

\bibitem{Katz:1984}
Katz S, Lebowitz J~L and Spohn H, {\em Nonequilibrium steady states of
  stochastic lattice gas models of fast ionic conductors\/},  1984 {\em J.
  Stat. Phys.\/} {\bf 34} 497--537

\bibitem{Schmittmann:1998}
Schmittmann B and Zia R~K~P, {\em Driven diffusive systems. An introduction and
  recent developments\/},  1998 {\em Phys. Rep.\/} {\bf 301} 45--64

\bibitem{Marro:1999}
Marro J and Dickman R 1999,  {\em Nonequilibrium Phase Transitions in Lattice
  Models\/} (Cambridge University Press)

\bibitem{SchmittmannBook:1995}
Schmittmann B and Zia R~K~P 1995,  {\em Statistical Mechanics of Driven
  Diffusive Systems\/} (Academic Press)

\bibitem{Caracciolo:2004}
Caracciolo S, Gambassi A, Hakim V, Gubinelli M and Pelissetto A, {\em
  Finite-Size Scaling in the Driven Lattice Gas\/},  2004 {\em J. Stat.
  Phys.\/} {\bf 115} 281--322

\bibitem{Schmittmann:1994}
Bassler K~E and Schmittmann B, {\em Renormalization-group study of a hybrid
  driven diffusive system\/},  1994 {\em Phys. Rev. E\/} {\bf 49} 3614--3618

\bibitem{del_Campo:2015}
del Campo A and Sengupta K, {\em Controlling quantum critical dynamics of
  isolated systems\/},  2015 {\em Eur. Phys. J. Spec. Top.\/} {\bf 224}
  189--203

\bibitem{Karaman:2012}
{Karaman} S and {Frazzoli} E 2012,  {\em High-speed flight in an ergodic
  forest\/}, {\em 2012 IEEE International Conference on Robotics and
  Automation\/} pp 2899--2906

\bibitem{Priyanka:2020}
Priyanka, T\"auber U~C and Pleimling M, {\em Feedback control of surface
  roughness in a one-dimensional Kardar-Parisi-Zhang growth process\/},  2020
  {\em Phys. Rev. E\/} {\bf 101} 022101

\bibitem{Li:2012}
Li L and Pleimling M, {\em Formation of nonequilibrium modulated phases under
  local energy input\/},  2012 {\em EPL\/} {\bf 98} 30004

\bibitem{Pleimling:2014}
Borchers N, Pleimling M and Zia R~K~P, {\em Nonequilibrium statistical
  mechanics of a two-temperature Ising ring with conserved dynamics\/},  2014
  {\em Phys. Rev. E\/} {\bf 90} 062113

\bibitem{Colangeli:2018}
Colangeli M, Giardin\`a C, Giberti C and Vernia C, {\em Nonequilibrium
  two-dimensional Ising model with stationary uphill diffusion\/},  2018 {\em
  Phys. Rev. E\/} {\bf 97} 030103

\bibitem{Sadhu:2012}
Sadhu T, Shapira Z and Mukamel D, {\em Interface Phase Transition Induced by a
  Driven Line in Two Dimensions\/},  2012 {\em Phys. Rev. Lett.\/} {\bf 109}
  130601

\bibitem{Dickman:2016}
Dickman R, {\em Phase coexistence far from equilibrium\/},  2016 {\em New J.
  Phys.\/} {\bf 18} 043034

\bibitem{Praest:2000}
Pr{\ae}stgaard E~L, Schmittmann B and Zia R~K~P, {\em A lattice gas coupled to
  two thermal reservoirs: Monte Carlo and field theoretic studies\/},  2000
  {\em Eur. Phys. J. B\/} {\bf 18} 675–695

\bibitem{Janssen:1986}
Janssen H~K and Schmittmann B, {\em Field Theory of Long Time Behaviour in
  Driven Diffusive Systems\/},  1986 {\em Z. Phys. B\/} {\bf 63} 517--520

\bibitem{Daquila:2011}
Daquila G~L and T{\"a}uber U~C, {\em Slow relaxation and aging kinetics for the
  driven lattice gas\/},  2011 {\em Phys. Rev. E\/} {\bf 83} 051107

\bibitem{Mukhamadiarov:2019}
Mukhamadiarov R~I, Priyanka and T{\"a}uber U~C, {\em Transverse temperature
  interfaces in Katz-Lebowitz-Spohn driven lattice gas\/},  2019 {\em Phys.
  Rev. E\/} {\bf 100} 062122

\bibitem{Krug:1991}
Krug J, {\em Boundary-induced phase transitions in driven diffusive systems\/},
   1991 {\em Phys. Rev. Lett.\/} {\bf 67} 1882--1885

\bibitem{Derrida:1992}
Derrida B, Domany E and Mukamel D, {\em An Exact Solution of a One-Dimensional
  Asymmetric Exclusion Model with Open Boundaries\/},  1992 {\em J. Stat.
  Phys.\/} {\bf 69} 667--687

\bibitem{Schutz:1995}
Sch{\"u}tz G and Domany E, {\em Phase Transitions in an Exactly Soluble
  One-Dimensional Exclusion Process\/},  1993 {\em J. Stat. Phys.\/} {\bf 72}
  277--296

\bibitem{Blythe:2007}
Blythe R~A and Evans M~R, {\em Nonequilibrium steady states of matrix-product
  form: a solver's guide\/},  2007 {\em J. Phys. A: Math. Theor.\/} {\bf 40}
  R333--R441

\bibitem{Spitzer:1970}
Spitzer F, {\em Interaction of Markov processes\/},  1970 {\em Adv. Math.\/}
  {\bf 5} 246--290

\bibitem{Liggett:1985}
Liggett T~M 1985,  {\em Interacting Particle Systems\/} (Springer-Verlag Berlin
  Heidelberg)

\bibitem{Leung:1986}
Leung K and Cardy J~L, {\em Field theory of critical behavior in a driven
  diffusive system\/},  1986 {\em J. Stat. Phys.\/} {\bf 44} 497--537

\bibitem{Daquila:2012}
Daquila G~L and T{\"a}uber U~C, {\em Nonequilibrium Relaxation and Critical
  Aging for Driven Ising Lattice Gases\/},  2012 {\em Phys. Rev. Lett.\/} {\bf
  108} 110602

\bibitem{Schmittmann:1995}
Schmittmann B and Zia R~K~P, {\em On singularities in the disordered phase of a
  driven diffusive system\/},  1995 {\em Z. Phys. B\/} {\bf 97} 327--332

\bibitem{Anderson:1987}
Anderson C~R and Greengard C (eds) 1987,  {\em Vortex Methods\/}
  (Springer-Verlag Berlin Heidelberg)

\bibitem{Tauber_2014}
T{\"a}uber U~C 2014,  {\em Critical Dynamics: A Field Theory Approach to
  Equilibrium and Non-Equilibrium Scaling Behavior\/} (Cambridge University
  Press)

\bibitem{Jiang:2004}
Jiang D~Q, Qian M and Qian M~P 2004,  {\em Mathematical Theory of
  Nonequilibrium Steady States\/} (Springer-Verlag Berlin Heidelberg)

\bibitem{Spohn:1991}
Spohn H 1991,  {\em Large Scale Dynamics of Interacting Particles\/}
  (Springer-Verlag Berlin Heidelberg)

\bibitem{Cates:2020}
Caballero F and Cates M~E, {\em Stealth Entropy Production in Active Field
  Theories near Ising Critical Points\/},  2020 {\em Phys. Rev. Lett.\/} {\bf
  124} 240604

\end{thebibliography}

\end{document}